\documentclass[useAMS,usenatbib]{mn2e}
\usepackage{graphicx}
\usepackage{natbib}
\usepackage{color}
\newcommand{\epsB}{\epsilon_{B,-1}}
\newcommand{\epse}{\epsilon_{e,-1}}
\newcommand{\G}{\Gamma}
\newcommand{\Go}{\Gamma_0}
\newcommand{\bo}{\beta_0}
\newcommand{\be}{\beta}
\newcommand{\g}{\gamma}
\newcommand{\nuo}{\nu_{obs}}
\newcommand{\td}{t_{dec}}
\newcommand{\msun}{$m_\odot$}
\newcommand{\apj}{Ap. J.}
\newcommand{\apjl}{Ap. J. Lett.}
\newcommand{\apjs}{Ap. J. Supp.}
\newcommand{\physrep}{Phys. Rep.}
\newcommand{\mnras}{Mon. not. RAS.}

\newcommand{\Msun}{$m_\odot$ }
\newcommand{\aap}{Astron. \& Astrophys.}
\newcommand{\prd}{Phys. Rev. D.}
\newcommand{\nat}{Nature}
\newcommand{\pasp}{Publications of the Astronomical Society of the Pacific}
\def\lesssim{\mathrel{\hbox{\rlap{\hbox{\lower4pt\hbox{$\sim$}}}\hbox{$<$}}}}
\def\gtrsim{\mathrel{\hbox{\rlap{\hbox{\lower4pt\hbox{$\sim$}}}\hbox{$>$}}}}

\title[The Electromagnetic Signals of  Compact Binary  Mergers]
{{\bf The Electromagnetic Signals of  Compact Binary  Mergers }}

\author[T. Piran, et al.]{ Tsvi Piran$^{1}$,
Ehud Nakar$^{2}$ 
and
Stephan Rosswog$^{3,4,5}$\\
$^{1}$Racah Institute of Physics, The Hebrew University, Jerusalem 91904, Israel\\
$^2$Raymond and Beverly Sackler School of Physics \& Astronomy, Tel Aviv University,
Tel Aviv 69978, Israel\\
$^3$School of Engineering and Science, Jacobs University Bremen, Germany\\
$^4$TASC, Department of Astronomy and Astrophysics, University of California, Santa Cruz, CA 95064\\
$^5$Astronomy and Oskar Klein Centre, Stockholm University, AlbaNova, SE-10691 Stockholm, Sweden}

\begin{document}

\maketitle
\begin{abstract}
Compact binary mergers  are prime sources of gravitational waves
(GWs), targeted by current and next generation detectors. The
question ``what is the observable electromagnetic (EM) signature of
a compact binary  merger?'' is an intriguing one with crucial
consequences to the quest for gravitational waves. We present a
large set of numerical simulations that focus on the electromagnetic
signals that emerge from the dynamically ejected sub-relativistic
material. These outflows produce on a time scale of a day macronovae
- short-lived  IR to UV  signals powered by radioactive decay.  Like
in regular supernovae the interaction of this outflow with the
surrounding matter inevitably leads to a long-lasting remnant.  We
calculate the expected radio signals of these remnants  on
time scales longer than a year, when the sub-relativistic ejecta
dominate the emission.  We discuss their detectability in 1.4 GHz
and 150 MHz and compare it with an updated estimate of the
detectability of short GRBs' orphan afterglows  (which are
produced by a different component of this outflow). We find that
mergers with characteristics similar to those of the Galactic
neutron star binary population (similar masses and typical
circum-merger Galactic disk density of $\sim 1 {\rm~cm^{-3}}$)  that
take place at the detection horizon of advanced GW detectors (300
Mpc) yield 1.4 GHz [150 MHz] signals of $\sim 50$ [300] $\mu$Jy, for
several years. The signal on time scales of weeks, is dominated by
the mildly and/or ultra-relativistic outflow, which is not accounted
for by our simulations, and is expected to be even brighter.
Upcoming all sky surveys are expected to detect a few dozen, and
possibly more, merger remnants at any given time thereby providing
 robust lower limits to the mergers rate  even before the advanced GW detectors
become operational. The macronovae signals from the same distance
peak in the  IR to UV range at an observed magnitude that may be
as bright as 22-23 about 10 hours after the merger but dimmer, redder and longer if the opacity is larger. 
\end{abstract}

\section{Introduction}
\label{sec:introduction} Compact binary  (neutron star - neutron
star, ns$^2$, or black hole - neutron star, nsbh) mergers are  prime
sources of gravitational radiation. The GW detectors LIGO
\citep{LIGO09}, Virgo \citep{Virgo08} and GEO600 \citep{GEO2008} are
designed to optimally detect merger signals. These detectors have
been operational intermittently during the last few years reaching
their nominal design sensitivity
\citep{LIGO09a,LIGOVirgo2010,LIGOVirgo2010a} with  detection
horizons of   a few dozen Mpc for ns$^2$ and almost a hundred Mpc
for nsbh mergers (the LIGO - Virgo collaboration adopts an optimal
canonical distance of 33/70Mpc;  \citealt{LIGORate2010}). Both LIGO
and Virgo are being upgraded now and by the end of 2015 are expected
to be operational at  sensitivities $\sim 10-15$ times greater than
the initial LIGO \citep{AdvancedLIGO2009}, reaching a detection
horizon of a few hundred Mpc for ns$^2$ mergers and about a Gpc for
nsbh mergers (445/927 Mpc are adopted by the LIGO-Virgo
collaboration as canonical values; \citealt{LIGORate2010}).

Understanding the observable EM signature of compact binary mergers
has several observational implications. First, once the detectors
are operational it is likely that the first detection of a GW signal
will be around or even below threshold. Detection of an accompanying
EM signal will confirm the discovery, thereby increasing
significantly the sensitivity of GW detectors
\citep{KP93,hughes03,dalal06,arun09}. Second, the physics that can
be learned from observations of a merger event through different
glasses is much greater than what we can learn through EM or GW
observations alone. Finally, even before the detectors are
operational, the detection of EM signatures will enable us to
determine the merger rates\footnote{The current rate constraints on
compact binary mergers are rather loose. The last LIGO and Virgo
runs provided only weak upper limits on the merger rates: $1.4
\times 10^4$ Myr$^{-1}$($10^{10}L_\odot)^{-1}$ corresponding to
$\sim 2 \times 10^5$  Gpc$^{-3}$ yr$^{-1}$ for ns$^2$ and 3600
Myr$^{-1}$ ($10^{10}L_\odot)^{-1}$  ($\sim 5 \times 10^4 {\rm
~Gpc^{-3}~yr^{-1}}$) for nsbh \citep{LIGO09a}. Estimates based on
the observed binary pulsars in the Galaxy are highly uncertain, with
values ranging from $ 20 - 2 \times 10^4  {\rm ~Gpc^{-3}~yr^{-1}}$
\citep{Phinney91,NPS91,KalEtal04,KalEtal04a,LIGORate2010}. There are
no direct estimates of nsbh mergers, as no such system has ever been
observed, and here one has to rely only on a  parameter dependent
population synthesis \citep[e.g.,][]{BKR+08,MO10}.}, an issue of
utmost importance for the design and the operation policy of the
advanced detectors.


The electromagnetic signal that is often considered as the most
promising counterpart to gravitational waves is that of short
Gamma-Ray Bursts (GRBs), which are thought to arise from compact
binary merger events \citep{Eichler89}. The estimated rate of short
GRBs is indeed comparable to binary pulsar estimates
\citep{GP06,NGF06,GS09}. However, while appealing, the association
is not proven yet \citep{Nakar07}, and even if there is an
association, short GRBs are observed only if their relativistic jets
points towards us. If short GRBs are binary mergers then their
observed rate, $\sim 10 {\rm ~Gpc^{-3}~yr^{-1}}$, provides a lower
limit to the merger rate. The true rate is higher and it depends on
the poorly constrained beaming angle, which results in an
uncertainty of almost two orders of magnitude. While a GRB that is
observed off-axis is undetectable in gamma-rays, it produces a
long-lasting radio ``orphan'' afterglow, which may be detectable
\citep{Rhoads97,Waxman+98,Frail+00,Levinson+02,Gal-Yam+06}. A key
point in estimating the detectability of GRB orphan afterglows is
that the well constrained observables are the  {\it isotropic}
equivalent energy of the flow and the rate of bursts that point
towards earth. However, the detectability of the orphan afterglows
depends only on the {\it total} energy and  {\it true} rate, namely
on the poorly constraint jet beaming angle. \cite{Levinson+02} have
shown that while narrower beaming increases the true rate it reduces
the total energy, and altogether reduces the detectability of radio
orphan afterglows. This counterintuitive result makes the
detectability of late emission from a decelerating jet, which
produced a GRB when it was still ultra-relativistic, less promising.

Regardless of the amount of ultra-relativistic outflow that is
launched by compact binary mergers, and of whether they produce
short GRBs or not, mergers do launch energetic sub-relativistic and
mildly-relativistic outflows
\citep[e.g.,][]{Rosswog+99,RJ01,Rosswog05,RP07,yst08,RBGLF10},
unless the equation of state at supra-nuclear densities is extremely
soft \citep{RDTP00}. Recently, \cite{NP11} showed that the
interaction of these outflows with the surrounding matter will
inevitably produce radio counterparts. An additional source of
electromagnetic signal was suggested by \cite{LP98}, which argue
that the freshly synthesized, radioactive elements in the ejected
debris from the merger will drive a short-lived supernova-like event
often referred to as ``macronova'' \citep{Kulkarni05}.
\cite{Metzger+10}  find that if $0.01 {\rm M_\odot}$ is ejected then
the optical emission from a merger at $300$ Mpc peaks after $\sim 1$
day at $m_V \approx 23$ mag. For a recent discussion of the
detectability of the various electromagnetic counterparts of GW
sources see \cite{MetzgerBerger12}.

In addition to electromagnetic signals, there will be a strong
($\sim 10^{53}$ erg) burst of $\sim 10$ MeV neutrinos, similar to
what is produced by a core-collapse supernova. But since compact
binary mergers are orders of magnitude rarer than supernova events,
the chances of detecting neutrinos from a cosmological merger event
are essentially zero.

In a companion paper (\citealt{RPN12}; in the following called
``paper I'') we have investigated to which extent dynamical
collisions as they occur, for example, in a globular cluster are
different from a gravitational wave driven compact binary merger. In
this paper we concentrate entirely on binary mergers and we
systematically explore the neutron star binary parameter space in a
large set of simulations. We use numerical simulations of the merger
process to find the properties of the dynamically ejected outflow
for different masses of the coalescing compact stars. We then take
the resulting ejecta profiles and calculate the electromagnetic
transients (i.e., radio flares and macronovae) that are related to
the dynamical ejecta of ns$^2$ and nsbh binaries.

Our study does not account for other types of outflows such as
neutrino-driven winds (which yield moderate velocities of $\sim 0.1$
c; \citealt{dessart09}) or mildly and/or ultra-relativistic outflows
that may emerge from close to the compact object at the center of
the merger. Since only the sub-relativistic dynamically ejected
material is explored we restrict the light curve calculation to time
scales of a year and longer, when the sub-relativistic component
dominates the emission. On shorter time scales of weeks and months
the mildly relativistic component dominates. Since the radio
luminosity depends strongly on the outflow velocity, emission on
time scales of weeks and months will be brighter than the one that
we find here even if the mildly relativistic component carries a
small fraction out of the total outflow energy (see \citealt{NP11}
for details). The radio flares depend sensitively on the surrounding
ISM density. We focus here on physical parameters found in known
Galactic ns binaries. Since all known binaries reside in the
Galactic disk, we consider a uniform density of 1 cm$^{-3}$
\cite{Draine11} as the most likely circum-merger
environment\footnote{ Ns$^2$ binaries are in random locations in
the Galactic disk. \cite{Draine11} estimates that 0.4 of the disk
volume has a density of about 0.6 cm$^{-3}$,  0.1 of the volume has
a much higher density while the rest has a low density.}.

Our new simulations, that focus on the ejecta also enable us to
revise the estimates of the macronovae light curves. These light
curves are determined  mostly by three ingredients: (i) the total
amount of the ejecta and their velocity structure that we calculate
here; (ii) the energy input from radioactive decay, for which we
use the most accurate estimates to date by Korobkin et al. (2012)
and (iii) the (poorly known) material opacity, taken from Metzger
et al. (2010).
At a finer level the light curve and the peak
luminosity depend also on the velocity distribution as the emission
from a given mass element moving with a specific velocity peaks when
this element becomes optically thin. We use the simulation results
to estimate this time scale and in this way we obtain macronovae
light curves for our different merger cases.

We begin (\S \ref{sec:simulations}) with  a brief discussion of our
numerical simulations (more details can be found in  paper I). We
focus in this discussion on the ejecta properties which are critical
both for radio flares and for macronovae. In  \S \ref{sec:Theory} we
first provide an analytic calculation of the radio emission
resulting from the interaction of single velocity ejecta  with the
surrounding interstellar matter (ISM). Then, we expand this solution
to an outflow with a distribution of velocities and we present  a
semi-analytic calculation to find the signal resulting from the
ejecta obtained in the simulations, which show a velocity range of
0.1-0.5 c. We discuss the observational implications for
detectability of merger remnants in \S \ref{sec:observations},
including a comparison to an updated estimate of short GRB orphan
afterglows detectability (\S 4.2). In \S \ref{sec:macronova} we
present a general theory of a macronova light curve for a case with
a velocity distribution within the ejecta. {We use in these
calculations new radioactive energy deposition  rates calculated
recently in a companion paper by \cite{korobkin12}. We then present
our detailed calculations of the expected IR to UV luminosity of
macronovae that arise from radioactive decay within the outflow. We
summarize our results and discuss their implications in \S
\ref{sec:conclusions}.

\begin{figure*}
\centerline{\includegraphics[angle=-90,width=22cm]{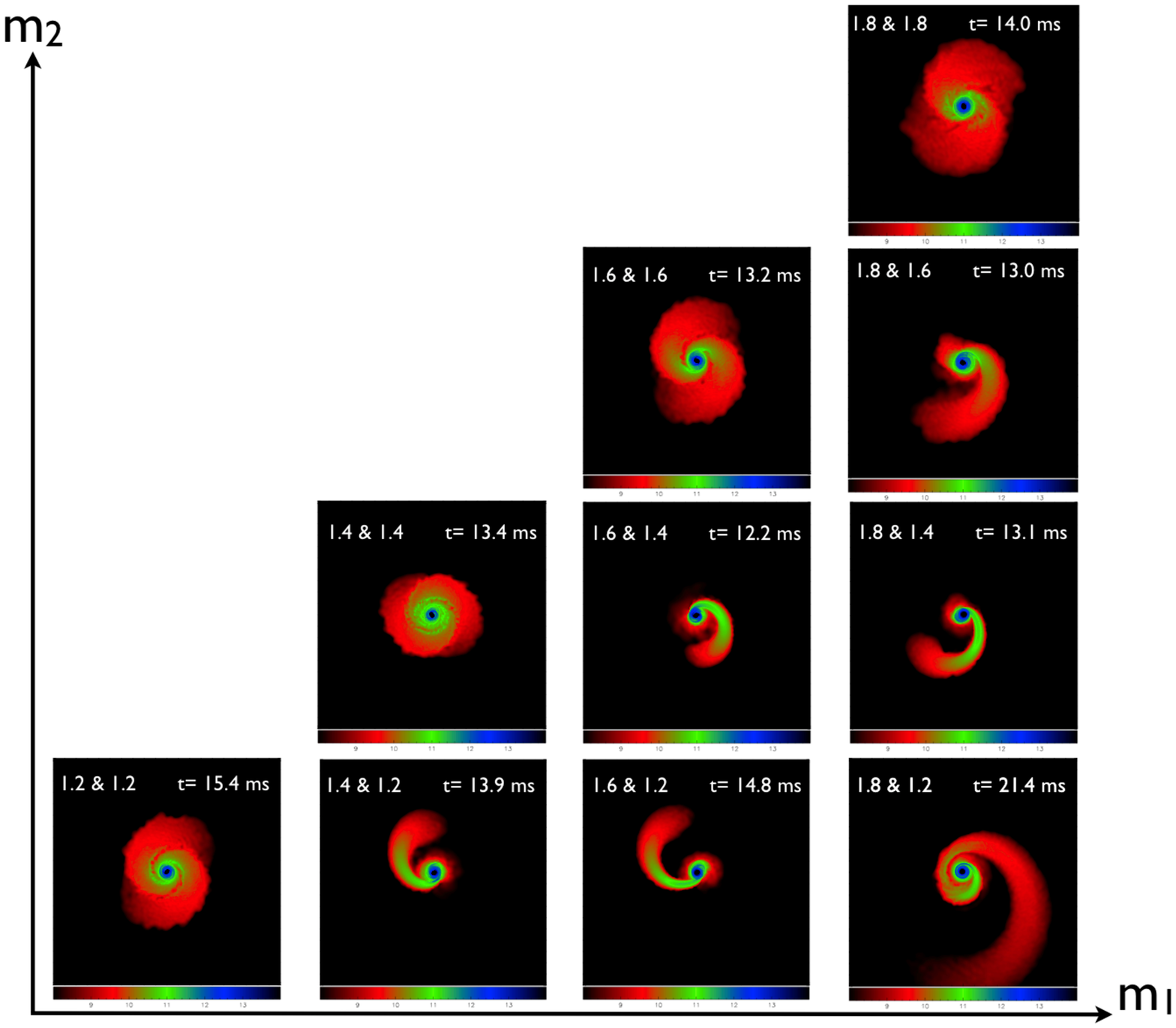}}
\caption{Density cuts through the orbital planes of all merger
remnants
         at the end of each simulation. Each snapshot shows a region of
         1000 km x 1000 km, color coded is the logarithm of mass density
         in g cm$^{-3}$, the annotations indicate the stellar masses (solar units)
         and the time of the snapshot.}
 \label{fig:XYstructures}

   \end{figure*}

\begin{figure*}
\centerline {
 \includegraphics[width=4cm,angle=-90]{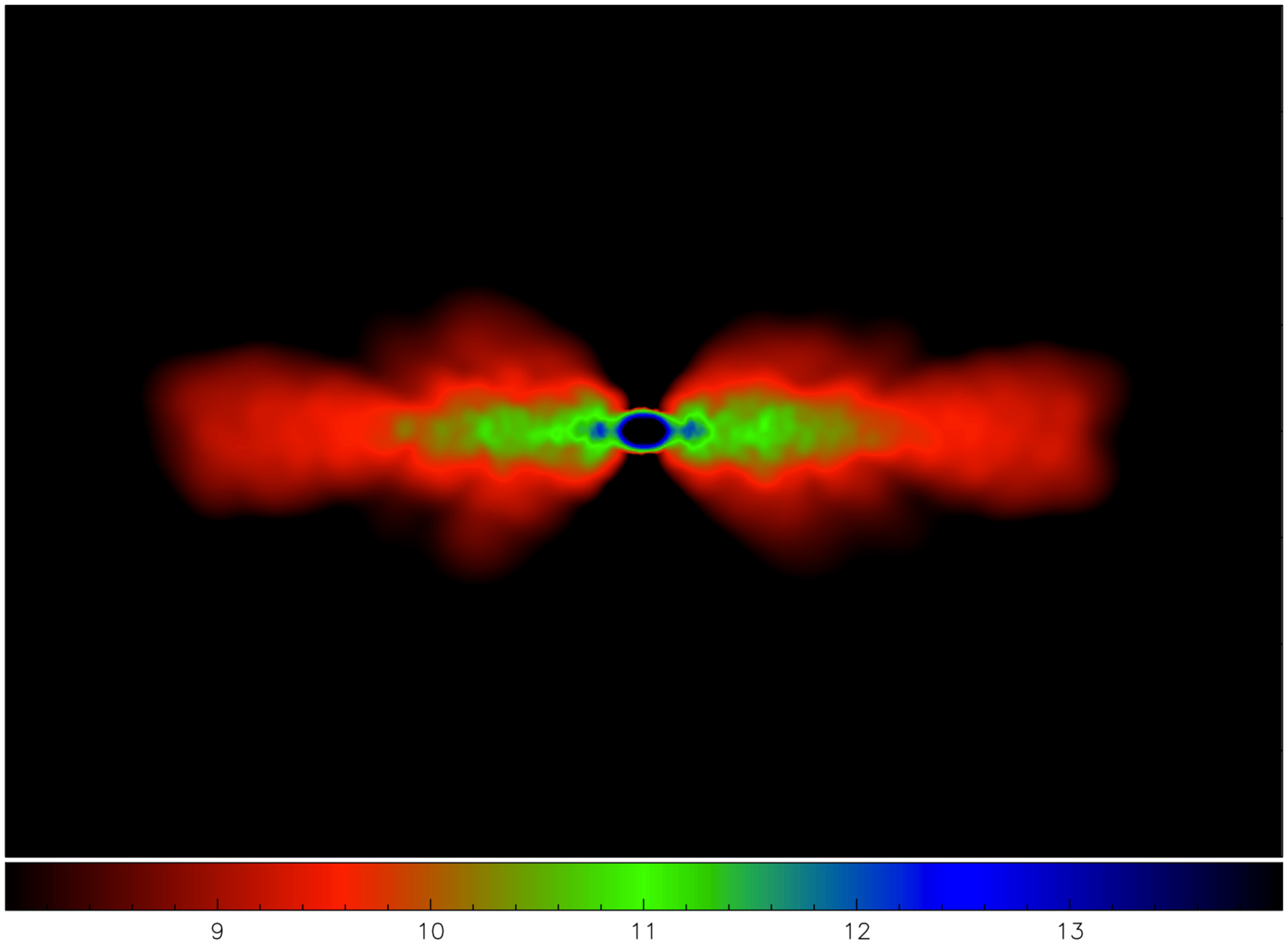}
 \includegraphics[width=4cm,angle=-90]{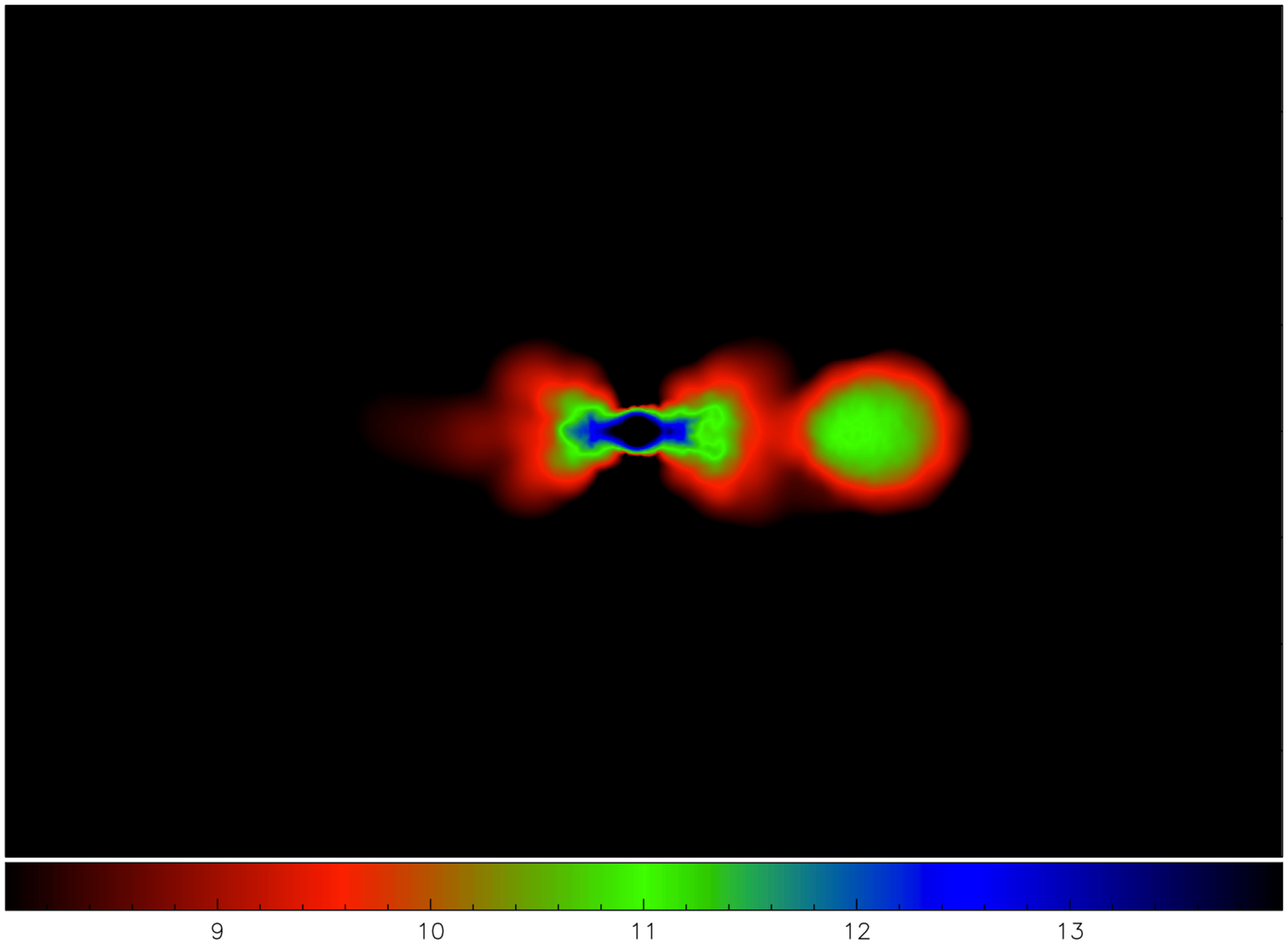}
\includegraphics[width=4cm,angle=-90]{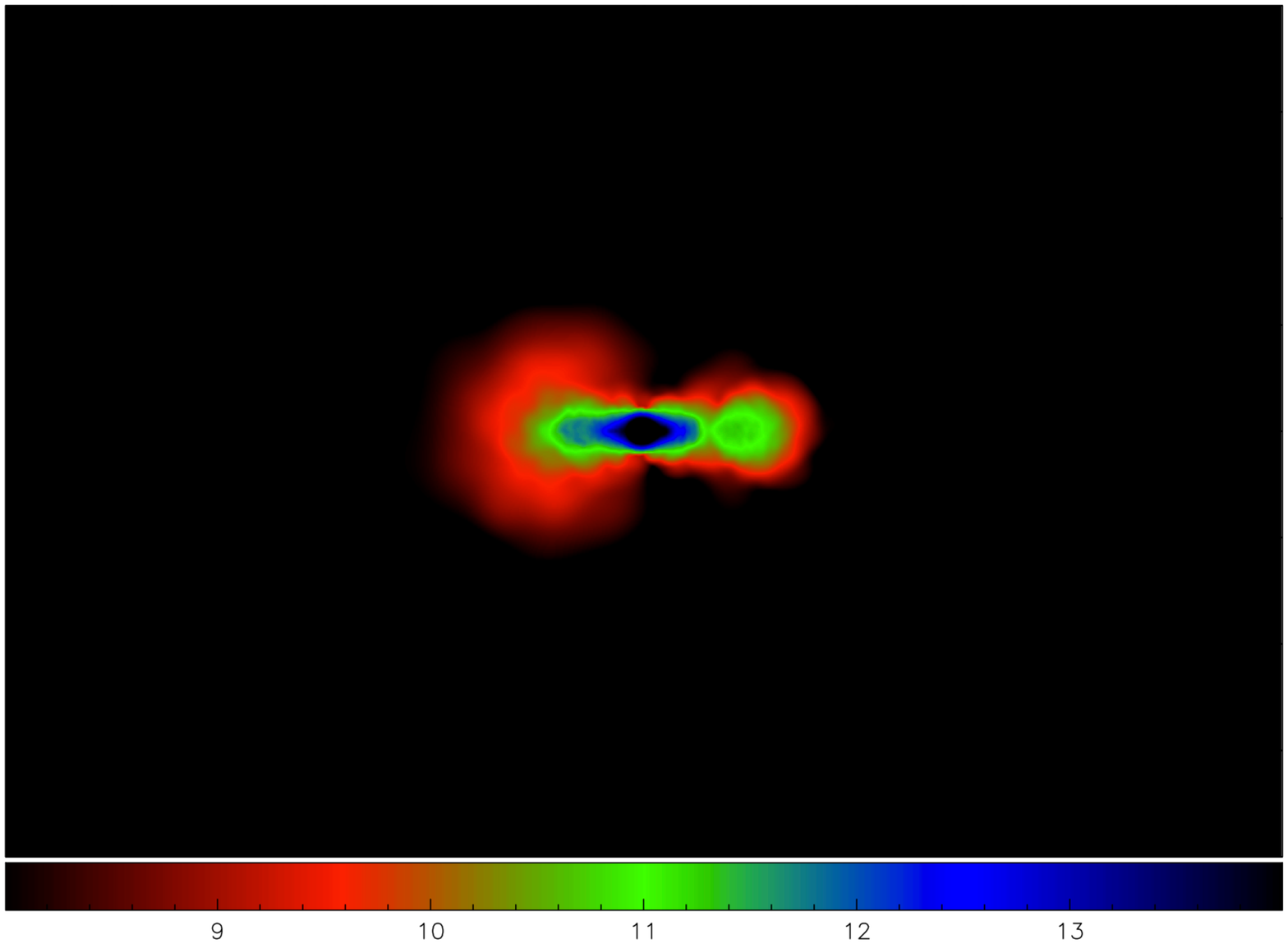}
}
   \caption{Density cuts (XZ-plane) at the last time slice of the simulations with
            2 x 1.4 \Msun  (left), 1.6 and 1.4 \Msun  (middle) and 1.8 and 1.4 \Msun  (right).
            The chosen cases correspond to the second line of panels (counted from below)
            in Fig.~\ref{fig:XYstructures}.}
   \label{fig:XZstructures}
   \end{figure*}

\section{Numerical simulations}
\label{sec:simulations} Neutron stars (ns) have long been thought to
be narrowly distributed around 1.35 \Msun \citep{thorsett99}. With
an increasing number of observed systems it has, however, turned out
that the mass range that is realized in nature is substantially
broader. For example, there is now ample support for neutron star
masses significantly larger than 1.5 \Msun (see for example the data
compilation in \citealt{lattimer10}). A broad peak around 1.5-1.7
\Msun has been found \citep{kiziltan10,valentim11} for neutron stars
with white dwarf companions. In addition, there may be a low-mass
peak of $\sim$1.25 \Msun neutron stars that have been produced by
electron capture supernovae
\citep{podsiadlowski04,vandenheuvel04,schwab10}. PSR J1614-2230 with
a mass of 1.97 $\pm$ 0.04 \Msun \citep{demorest10} is nowadays
considered as a robust lower limit on the maximum neutron star mass,
but even neutron stars with considerably larger
masses are not implausible.\\

These findings are more than enough of a motivation for a broad scan
of the neutron star binary parameter space. We explore neutron star
masses between 1.0 to 2.0 \Msun in steps of 0.2 \msun. All our
neutron stars have a negligible initial spin, consistent with the
results of \cite{bildsten92} and \cite{kochanek92}. Our simulations
make use of the Smooth Particle Hydrodynamics (SPH) method, see
\cite{monaghan05} and \cite{rosswog09b} for recent reviews. Our code
is an updated version of the one that was used in earlier studies
\citep{rosswog02a,rosswog03a,rosswog03c,Rosswog05}. It uses  the
Shen et al. equation of state (EOS) \citep{shen98a,shen98b}, an
opacity-dependent multi-flavor neutrino leakage scheme
\citep{rosswog03a} and a time-dependent artificial viscosity
treatment, see \cite{rosswog00,rosswog08b} for details. In all nsbh
simulations Newtonian gravity was employed and the black hole was
vested with an absorbing boundary at the Schwarzschild radius. All
simulations used a simple gravitational wave emission backreaction
force \citep{davies94}. The performed simulations complement those
that have been presented in Paper I. For completeness, we have also
performed two simulations of neutron star black hole binary systems.
Since we found in earlier studies \citep{rosswog04} cases of
long-lived, episodic mass transfer, we started the two cases with
lower numerical resolution to be able to follow them until the
neutron star is completely disrupted\footnote{Phases of stable
mass transfer are not necessarily restricted to simulations that use
the approximation of Newtonian gravity. Relativistic binary
simulations with small mass ratios and large bh spin parameters may
be particularly prone to a stable mass transfer phase, see \cite{shibata11}
for a further discussion.}. Our results have turned out to be very
robust with respect to the numerical resolution, therefore the
reduced numerical resolution is not a concern for the purpose of our
study.  The system parameters and some key properties of the ejecta are summarized in Table ~\ref{tab:cases}.\\
In all investigated cases we find $\sim10^{-2}$ \Msun of unbound material ($E_{\rm kin} + E_{\rm pot} > 0$).
A deviation of the mass ratio from unity has the tendency to enhance the amount of ejected mass, we find
in contrast no clear tendency with the total system mass, see column six in Table ~\ref{tab:cases}. In all
cases the ejecta velocities are below 0.5 c, their mass-averaged values are given in column seven, they
are typically close to 0.13 c.\\
\\
\begin{table*}
 \centering
 \begin{minipage}{140mm}
  \caption{Overview over the performed simulations}
  \begin{tabular}{@{}cccclccccl@{}}
  \hline
   Run   &  $m_1$ [\msun] & $m_2$ [\msun] & $N_{\rm SPH}\; [10^6]$ & $t_{\rm end}$ [ms] & $m_{\rm ej}$ [\msun]& $\langle v_{\rm esc}\rangle$ & $E_{\rm kin} [10^{50} {\rm erg}]$ \\
\hline
\\
   1     &   1.0          & 1.0        & 1.0    &  15.3   & $ 7.64 \times 10^{-3}$ & 0.10   & 1.0 \\
   2     &   1.2          & 1.0        & 1.0    &  15.3   & $ 2.50 \times 10^{-2}$ & 0.11   & 3.4\\
   3     &   1.4          & 1.0        & 1.0    &  16.5   & $ 2.91\times 10^{-2}$  & 0.13   & 4.8\\
   4     &   1.6          & 1.0        & 1.0    &   31.3  & $ 3.06\times 10^{-2}$  & 0.13   & 5.3\\
   5     &   1.8          & 1.0        & 1.0    &   30.4 *  & $>1.64 \times 10^{-2}$& 0.13& 3.2 \\
   6     &   2.0          & 1.0        & 0.6    &   18.8 * & $>2.39\times 10^{-2}$ & 0.16& 6.0 \\
\\
   7    &   1.2          & 1.2         & 1.0    &  15.4  & $ 1.68 \times 10^{-2}$ & 0.11 & 2.3 \\
   8    &   1.4          & 1.2         & 1.0    &  13.9  & $ 2.12 \times 10^{-2}$ & 0.12 & 3.2\\
   9    &   1.6          & 1.2         & 1.0    &  14.8  & $ 3.33 \times 10^{-2}$ & 0.13 & 6.2 \\
   10  &   1.8          & 1.2         & 1.0     &  21.4  & $ 3.44 \times 10^{-2}$ & 0.14  & 7.0\\
   11  &   2.0          & 1.2         & 0.6     &  15.1 *& $>2.95 \times 10^{-2}$& 0.14&6.0& \\
\\
  12    &   1.4         & 1.4         & 1.0     &  13.4  & $1.28 \times 10^{-2}$  & 0.10 & 1.6\\
   13   &   1.6         & 1.4         & 1.0     &  12.2  & $2.36\times 10^{-2}$   & 0.12 & 4.0\\
   14   &   1.8         & 1.4         & 1.0     &  13.1  & $3.84\times 10^{-2}$   & 0.14 & 7.6\\
   15   &   2.0         & 1.4         & 0.6     &  15.0  & $3.89\times 10^{-2}$   & 0.15 & 8.7\\
\\
   16   &   1.6         & 1.6          & 1.0    &  13.2  & $1.97 \times 10^{-2}$ & 0.11 & 2.9\\
   17   &   1.8         & 1.6          & 1.0    &  13.0  & $1.67 \times 10^{-2}$ & 0.12 & 2.7\\
   18   &   2.0         & 1.6          & 0.6    &  12.4  & $3.79\times 10^{-2}$  & 0.14 & 7.6\\
  \\
  19   &   1.8         & 1.8          & 1.0     &  14.0  & $1.50 \times 10^{-2}$ & 0.12 & 2.8\\
  20   &   2.0          & 1.8         & 0.6     &  11.0  & $1.99 \times 10^{-2}$ & 0.13 & 3.7\\
  \\
  21   &   2.0         & 2.0          & 0.2     &  21.4  &  $1.15\times 10^{-2}$ & 0.11 & 1.8\\
  \\
  22   &     5+      & 1.4          &    0.2     &   138.7 &     $2.38 \times 10^{-2}$ &  0.15  &        6.0   \\
  23   &   10+       & 1.4          &     0.2    &  139.3  &  $ 4.93 \times  10^{-2}$   &  0.18  &   18.2   \\
  \hline \hline
\label{tab:cases}
\end{tabular}
\\
$*$ The secondary is still orbiting at the end of the computations.\\
+ The primary is a black hole.\\
\end{minipage}
\end{table*}
Figs. \ref{fig:XYstructures} and \ref{fig:XZstructures} depict the
remnant structures in the orbital (XY) plane and perpendicular to it
(XZ) at the end of the simulations. Clearly, visible is the
sensitivity to deviations from a mass ratio of unity: even
differences of 15\% in the stellar masses lead to large asymmetries,
i.e. one pronounced tidal tail rather than a disk resulting from
initially two such tails. The debris matter cannot cool rapidly
enough, therefore it is puffed up, see Fig.~\ref{fig:XZstructures},
but at the end of the simulations far from spherical symmetry.
Although this may lead to some viewing angle dependencies, we assume
in the following for simplicity spherically symmetric outflows.\\
\begin{figure}
\centerline{\includegraphics[width=10cm]{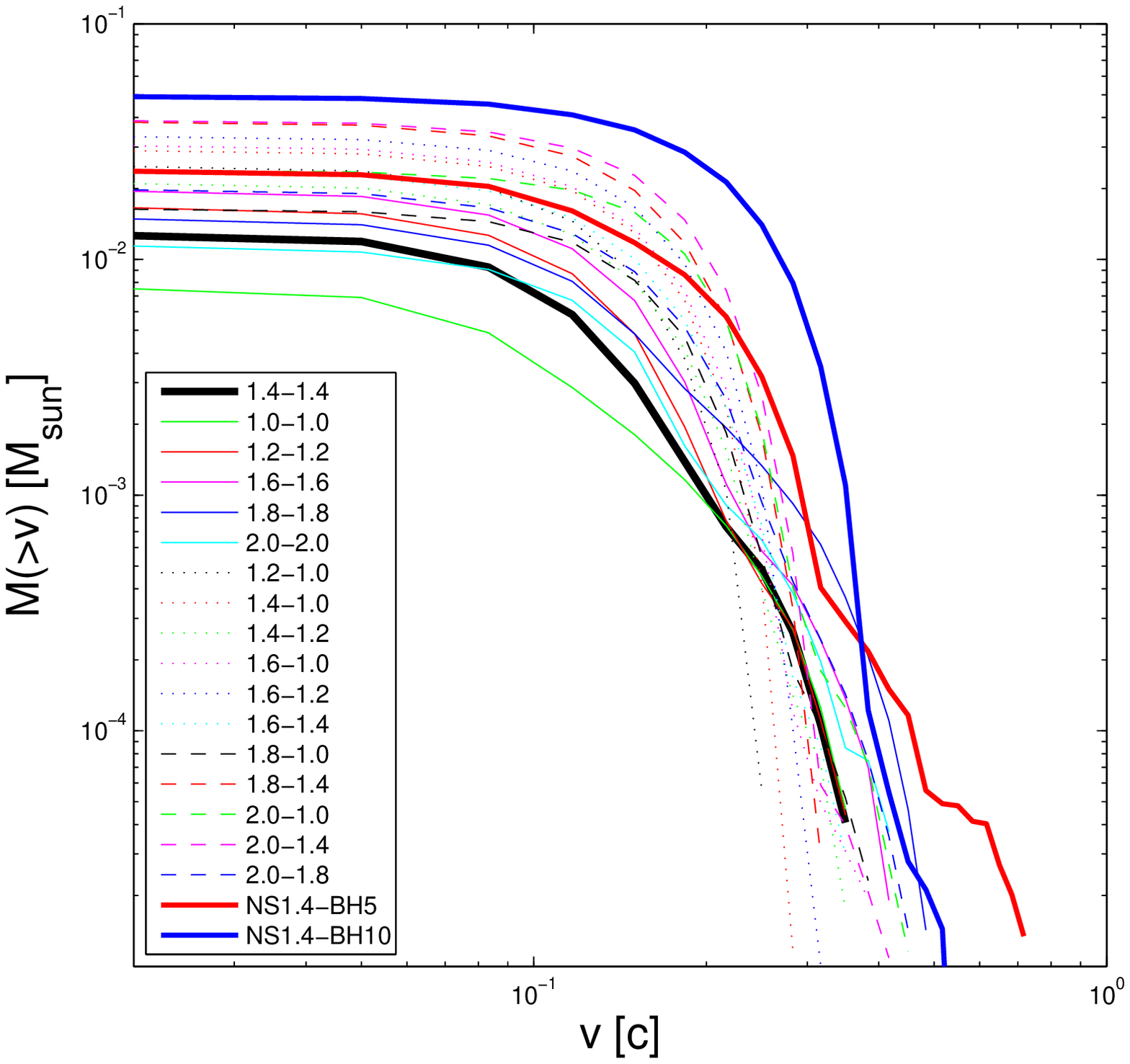}} \caption{Ejected mass with energy
above a given velocity for the different merger cases.}
\label{fig:Mv}
\end{figure}
Compact binary mergers inevitably eject mass in various forms:
\begin{itemize}
\item[a)] If compact binaries indeed power short GRBs, they
have to launch ultra-relativistic outflows with Lorentz factors
$>$100-1000. There are several mechanisms by which this could be
achieved (e.g., \citealt{BZ77,Hawley06,piran04,lee07,Nakar07})
\item[b)] Once the merger has happened, the hot remnant emits a few times
$10^{53}$ erg/s in $\sim 20$ MeV neutrinos
\citep{Eichler89,RJ01,rosswog03a,sekiguchi11}. The peak of the neutrino
emission is delayed with respect to the merger itself by the time it
takes to form a hot accretion torus, about 10 ms. At these huge
luminosities the neutrinos drive a strong baryonic wind of
$\dot{M}\sim 10^{-3}$ \msun/s and $v\sim 0.1c$ \citep{dessart09},
preferentially in the direction of the rotation axis.
\item[c)] It is very likely that the outflows of process a) and b) interact
near the rotation axis/the inner disk. Such an interaction plausibly produces
moderately relativistic matter outflows near the jet-wind interface.
\item[d)] As the accretion disk evolves it spreads viscously until
dissipational heating and/or the recombination of nucleons into nuclei
unbind a large fraction of the late-time disk \citep{chen07,lee07,metzger08,beloborodov08}.
\item[e)] Gravitational torques dynamically eject matter directly at first contact
with velocities $> 0.1$ c, see Tab.~\ref{tab:cases}.
\end{itemize}
While all of the above mass loss processes undoubtedly occur, the
quantitative calculation of processes a) to d) is technically very
demanding and the relevant physical processes are not included in
the presented simulations.  We therefore do not consider here their
contribution to the electromagnetic signature (which will be most
important for the radio emission at early times - as discussed
later). Instead, we focus entirely on the signature of the dynamic
ejecta, which can be reliably calculated\footnote{Note, that the
presented ejecta amounts are numerically fully converged. They can,
however, depend on the included physics ingredients such as the
equation of state or the treatment of gravity (Newtonian vs. GR).}.
Their properties are entirely set during the first contact and by
the end of the simulation the unbound material is moving
ballistically, with fast moving ejecta ahead of slower one. The
ballistic motion ends only when the dynamical ejecta start being
decelerated by the ambient material months-years after the merger
(see below). The  dynamically ejected debris is expected to be the most
energetic of all the outflow components. As a result its velocity
profile is not expected to be significantly affected by interaction
with other outflow components and its profile at the end of the
numerical simulation provides a good approximation of the initial
conditions for the electromagnetic signal calculations.

\section{The radio signal from outflow-ISM interaction}
\label{sec:Theory} We begin by calculating  analytically the radio
signal resulting from a single velocity outflow and from an outflow
with a power-law velocity distribution. We begin in section \S \ref{sec:Theory.1}
by repeating (for completeness) and extending   the calculation discussed in the supplementary material of \cite{NP11}. 
We then use semi-analytical calculations to generalize the result to ejecta with
an arbitrary distribution of velocities.

\subsection{A single velocity outflow}
\label{sec:Theory.1}
Consider a spherical outflow with an energy $E$ and an initial
Lorentz factor $\Go$, with a corresponding velocity $c \beta_0$,
that propagates into a constant density, $n$, medium. If the outflow
is not ultra-relativistic, i.e., $\Go-1 \lesssim 1$  it propagates
at a constant velocity until, at $t_{dec}$, it reaches radius
$R_{dec}$, where it collects a comparable mass to its own \citep{NP11}:
\begin{equation}
    R_{dec} = \left( \frac{3E}{4\pi n m_p c^2 \bo^2} \right)^{1/3} \approx 10^{17}
    {\rm ~cm~} E_{49}^{1/3} n^{-1/3} \bo^{-2/3},
\end{equation}
and
\begin{equation}
    t_{dec}= \frac{R_{dec}}{c\beta_0} \approx 30 {\rm ~day~}
    E_{49}^{1/3} n^{-1/3} \bo^{-5/3} ,
    \label{eq:tdec}
\end{equation}
where we approximate $\Go-1 \approx \bo^2$ and ignore relativistic
effects. Here and in the following, unless stated otherwise, $q_x$
denotes the value of $q/10^x$ in c.g.s. units. At a radius
$R>R_{dec}$ the flow decelerates assuming the Sedov-Taylor
self-similar solution, so the outflow velocity can be approximated
as:
\begin{equation}\label{eq Gamma}
    \be = \bo \left\{\begin{array}{cc}
                   1 & R \leq R_{dec}~,  \\
                   \left({R}/{R_{dec}}\right)^{-3/2} & R > R_{dec} ~.
                \end{array} \right.
\end{equation}

If the outflow is collimated, highly relativistic and points away
from a generic observer, as will typically happen if  the mergers
produce short GRBs, the emission during the relativistic phase will
be suppressed by relativistic beaming. Observable emission is
produced only once the external shock decelerates to mildly
relativistic velocities and the blast-wave becomes quasi spherical.
This takes place when $\Gamma \approx 2$ namely at $R_{dec}(\bo =
1)$. From this radius the hydrodynamics and the radiation become
comparable to that of a spherical outflow with an initial Lorentz
factor $\G_0 \approx 2$. This behavior is the source of the late
radio GRB orphan afterglows \citep{Rhoads97,Levinson+02}. Our theory
is therefore applicable for the detectability of mildly and
non-relativistic outflows as well as  for radio orphan GRB
afterglows.

Emission from Newtonian and mildly relativistic shocks is observed
in radio SNe and late phases of GRB afterglows. These observations
are well explained by a theoretical model involving synchrotron
emission of shock accelerated electrons in an amplified magnetic
field.  The success of this model in explaining the detailed
observations of radio Ib/c SNe
\citep[e.g.,][]{Chevalier98,Soderberg+05,CF06} allows us to employ
the same microphysics here. Energy considerations show that both the
electrons and the magnetic field carry significant fractions of the
total internal energy of the shocked gas, $\epsilon_e \approx
\epsilon_B \sim 0.1$ . These values are consistent with those
inferred from late radio afterglows of long GRBs
\cite[e.g.,][]{Frail+00,Frail+05}. The observed spectra indicate
that the distribution of the accelerated electrons' Lorentz factors,
$\g$,  is a power-law $dN/d\g \propto \g^{-p}$ at $\g>\g_m$ where $p
\approx 2.1-2.5$ in mildly relativistic shocks (e.g., the radio
emission from GRB associated SNe and late GRB afterglows) and $p
\approx 2.5-3$ in Newtonian shocks (as seen in typical radio SNe;
\citealt[][and references therein]{Chevalier98}). The value of
$\g_m$ is not observed directly but it can be calculated based on
the total energy of the accelerated electrons, $\g_m =
\frac{p-2}{p-1} \frac{m_p}{m_e} \epsilon_e \be^2$.

The radio spectrum generated by the shock is determined by two
characteristic frequencies\footnote{The cooling frequency is
irrelevant in the radio.}. One is
\begin{equation}\label{eq num}
    \nu_m \approx 1 {\rm ~GHz~} n^{1/2} \epsB^{1/2} \epse^{2} \be^{5} ,
\label{eq:num}
\end{equation}
the typical synchrotron frequency of electrons with the typical
(also minimal) Lorentz factor $\g_m$. The other is $\nu_a$, the
synchrotron self-absorption frequency. We show below that since we
are interested in the maximal flux at a given observed frequency,
$\nu_a$ may play a role only if it is larger than $\nu_m$. Its value
in that case is
\begin{equation}\label{eq nua}
    \nu_a (>\nu_m) \approx 1 {\rm ~GHz~} R_{17}^\frac{2}{p+4}
    n^\frac{6+p}{2(p+4)} \epsB^\frac{2+p}{2(p+4)} \epse^\frac{2(p-1)}{p+4}
    \be^\frac{5p-2}{(p+4)} .
\label{eq:nua}
\end{equation}

\begin{figure}
\centerline{
  \includegraphics[width=8cm]{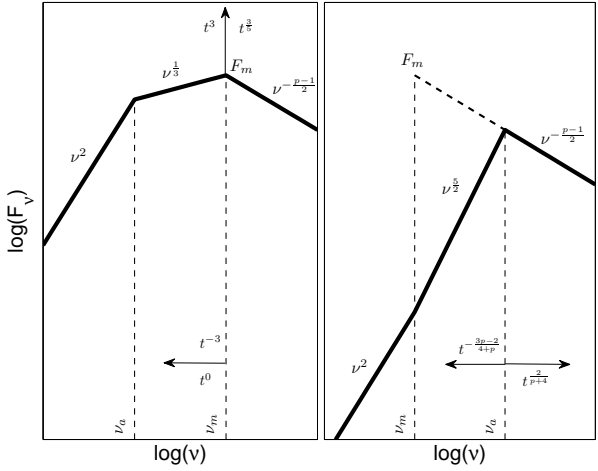}
}
  \caption{A sketch of the two possible spectra and the evolution of the characteristic flux, $F_m$, and frequencies, $\nu_a$ and $\nu_m$.
  The arrows show the temporal evolution of the characteristic values. The temporal dependence before $t_{dec}$ is
  noted below/to the left of the arrows while the temporal dependence after $t_{dec}$ is
  noted above/to the right of the arrows. Note that the evolution of $\nu_m$ and $F_m$, marked only in the left spectrum,
  is relevant for both spectra. The evolution of $\nu_a$, marked only in the right spectrum, is correct only when $\nu_m<\nu_a$ and is
  therefore relevant only in that spectrum.}\label{fig: spec}
\end{figure}

Fig. \ref{fig: spec} illustrates the two possible spectra,
depending on the order of $\nu_a$ and $\nu_m$. The flux at any
frequency can be found using these spectra and  the unabsorbed
synchrotron flux at $\nu_m$:
\begin{equation}\label{eq Fm}
    F_m \approx 0.5 {\rm ~mJy~}  R_{17}^3 n^{3/2} \epsB^{1/2}
    \be d_{27}^{-2} ,
\end{equation}
where $d$ is the distance to the source (we neglect any cosmological
effects). Note that this is the real flux at $\nu_m$ only if $\nu_a
< \nu_m$ (see Fig. \ref{fig: spec}).

As long as the shock is moving with a constant velocity  i.e., at $t<t_{dec}$, the
flux across the whole spectrum increases (see
equations \ref{eq num}-\ref{eq Fm}).
The flux evolution at later times depends on the spectrum at
$t_{dec}$, namely on
\begin{equation}\label{eq numdec}
    \nu_{m,dec} \equiv \nu_m(t_{dec}) \approx 1 {\rm ~GHz~} n^{1/2} \epsB^{1/2} \epse^{2}
    \bo^{5} ,
\end{equation}
 and if $\nu_{a,dec}>\nu_{m,dec}$ then possibly on
\begin{equation}\label{eq nuadec}
 \nu_{a,dec} \equiv  \nu_a(t_{dec})  \approx 1 {\rm ~GHz~} E_{49}^\frac{2}{3(4+p)} n^\frac{14+3p}{6(4+p)} \epsB^\frac{2+p}{2(4+p)} \epse^\frac{2(p-1)}{4+p}
    \bo^\frac{15p-10}{3(4+p)} .
\end{equation}
The flux at that time can be found using the unabsorbed synchrotron flux at $\nu_{m,dec}$:
\begin{equation}
    F_{m,dec} \approx 0.5 {\rm ~mJy~} E_{49} n^{1/2} \epsB^{1/2}
    \bo^{-1} d_{27}^{-2} .
\end{equation}

Consider now a given observed frequency $\nuo$. We are interested in
the light curve near the peak flux at this frequency. There are
three possible types of light curves near the peak corresponding to:
(i) $\nu_{m,dec},\nu_{a,dec}<\nuo$, (ii) $\nu_{eq}< \nuo <
\nu_{m,dec}< $ and (iii) $ \nuo < \nu_{eq}, \nu_{a,dec} $.  Where we
define
\begin{equation}
    \nu_{eq} = 1 {\rm ~GHz~} E_{49}^{1/7} n^{4/7} \epsB^{2/7}
    \epse^{-1/7}
\end{equation}
as the  frequency at which\footnote{Note that if $\nu_{m,dec} <
\nu_{a,dec}$ this equality will never take place. In that case
$\nu_{eq}$ is the frequency at which this equality would have
happened if $\Go$ would have been large enough (see Fig.
\ref{fig:cases})} $\nu_m=\nu_a$. In Fig. \ref{fig:cases} we show a
sketch of the time evolution of $\nu_a$ and $\nu_m$ and
the corresponding ranges of $\nuo$ in which each of the cases is
observed.

\begin{figure}
  \centerline{\includegraphics[width=8cm,angle=90]{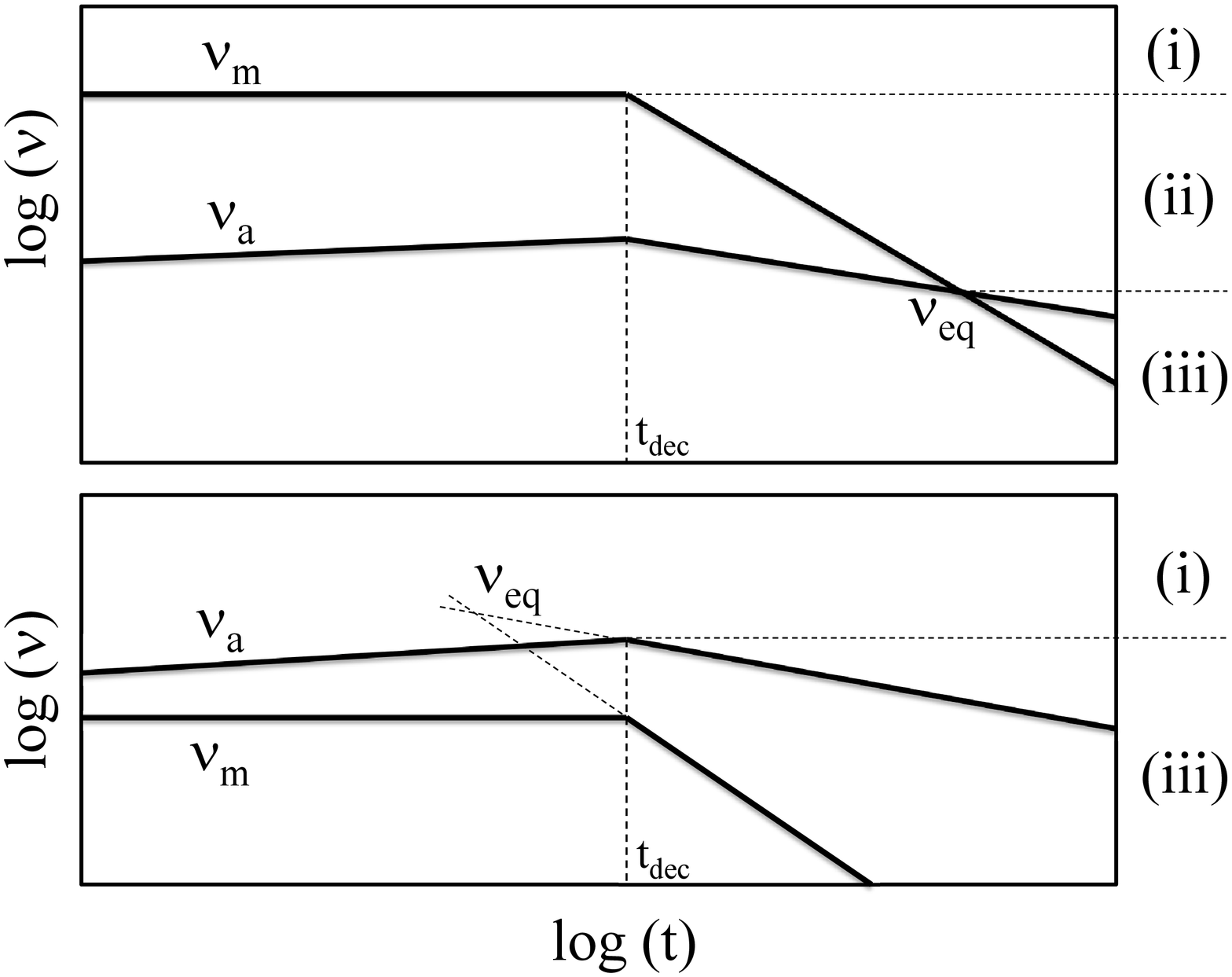}}
  \caption{A sketch of the time evolution of $\nu_a$ and $\nu_m$ in two
cases, $\nu_{a,dec}<\nu_{m,dec}$ (top) and $\nu_{a,dec}>\nu_{m,dec}$
(bottom). Also marked is the value of $\nu_{eq}$. The vertical
dashed line marks $\td$. The ranges of $\nuo$ at which each of the
cases is observed is separated by horizontal dashed lines and marked
on the right. Note that in the bottom panel $\nu_m$ and $\nu_a$ are
not crossing each other at $t>\td$ and only two types of light
curves, cases (i) and (iii), can be observed.}\label{fig:cases}
\end{figure}

To estimate the time and value of the peak flux we recall, that at
all frequencies  the flux increases until $t_{dec}$. In case (i),
$\nu_{m,dec},\nu_{a,dec}<\nuo$, the deceleration time, $\td$, is
also the time of the peak. The reason is that while $F_m$ increases,
$\nu_m$ decreases fast enough so that $F_{\nuo}$ decreases after
$t_{dec}$. Note that in that case $\nu_a$ plays no role since it
decreases after deceleration. Overall, in this case the flux peaks
at $t_{dec}$ and $F_{{\nuo},peak}= F_{m,dec}
(\nuo/\nu_{m,dec})^{-(p-1)/2}$.

In the two other cases, (ii) and (iii), $ \nuo < \nu_{m,dec}$ and/or
$\nuo<\nu_{a,dec}$  and the flux keeps rising at $t>\td$ until
$\nuo=\nu_{m}(t)$ or $\nuo=\nu_{a}(t)$, whichever comes last. To
find out which one of the two frequencies is it, we compare $\nuo$
with $\nu_{eq}$. At $t>t_{dec}$,  $\nu_m$ decreases faster than
$\nu_a$. Therefore in case (ii) where $\nu_{eq}<\nuo$,  the last
frequency to cross $\nuo$ is $\nu_m$ and the peak flux is observed
when $\nuo=\nu_m(t)$. In case (iii) where $\nuo<\nu_{eq}$, the last
frequency to cross $\nuo$ is $\nu_a$ and the peak flux is observed
when $\nuo=\nu_a(t)$. Now, it is straight forward to calculate the
peak flux, $F_{{\nuo},peak}$ and the time that it is observed,
$t_{peak}$, for different frequencies. It is also straight forward
to calculate the flux temporal evolution prior and after $t_{dec}$
using equations \ref{eq num}-\ref{eq Fm} and the relation $t \propto
R$ which holds at $t<t_{dec}$ and $\be \propto t^{-3/5}$ at
$t>t_{dec}$. The peak fluxes, the times of the peak and the temporal
evolution of the three different cases are summarized in table
\ref{table1}. The overall different light curves are depicted in
Fig. \ref{fig:lightcurves}

\begin{table*}
\begin{minipage}{140mm}
\begin{tabular}{|l|c|c|c|c|}
   \hline
  Case & $F_{{\nuo},peak}/F_{m,dec}$ & $t_{peak}/t_{dec}$ & $F_{\nuo}^\dag$ & $F_{\nuo}$ \\
  &  &  & $t<t_{peak}$ & $t_{peak}<t$ \\
  \hline\hline
  (i) $\nu_{m,dec},\nu_{a,dec}<\nuo$ & $
\left({\nuo}/{\nu_{m,dec}}\right)^{-\frac{p-1}{2}}$ & $1$ & $\propto
t^{3}$ & $\propto t^{-\frac{15p-21}{10}}$  \\
\hline
  (ii) $\nu_{eq}<\nuo<\nu_{m,dec}$ & $
\left({\nuo}/{\nu_{m,dec}}\right)^{-1/5}$ &
$({\nuo}/{\nu_{m,dec}})^{-1/3}$ & $\propto t^{\frac{8}{5}}$ &
$\propto t^{-\frac{15p-21}{10}}$  \\
\hline
  (iii) $\nuo<\nu_{eq},\nu_{a,dec}$ & $
~\nu_{m,dec}^\frac{p-1}{2}
~\nu_{a,dec}^{-\frac{3(p+4)(5p-7)}{10(3p-2)}}
~\nuo^\frac{(32p-47)}{5(3p-2)}$ &
$({\nuo}/{\nu_{a,dec}})^{-\frac{4+p}{3p-2}}$ & $\propto
t^{\frac{3}{2}}$ & $\propto t^{-\frac{15p-21}{10}}$  \\
  \hline
\end{tabular}
\caption{\label{table1} The observed flux before and after
$t_{peak}$ in the three different regimes. \newline $\dag$ The
temporal evolution only during the last power-law segment before
$t_{peak}$. At earlier times the temporal evolution may be
different.}
\end{minipage}
\end{table*}

\begin{figure}
  \hspace*{-2cm}\includegraphics[width=8cm,angle=-90]{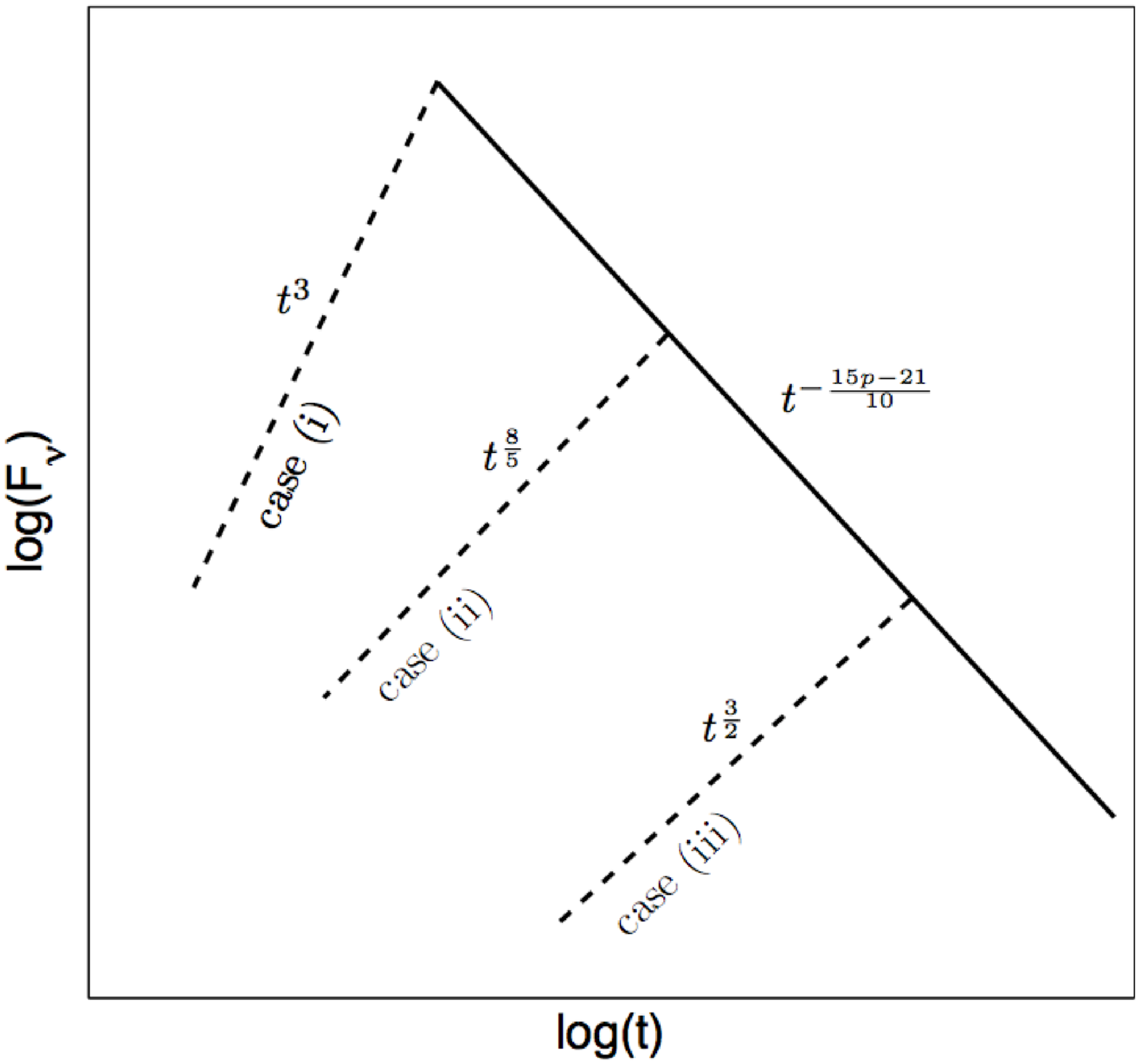}
  \caption{Schematic light curves of the three cases. The rising phase,
  marked in dashed line for each of the phases, is that of the last
  temporal power law segment before the peak. After the peak all cases
  show the same power-law decay. }\label{fig:lightcurves}
\end{figure}

The most sensitive radio facilities are at frequencies of $1.4$ GHz
and higher. Equations \ref{eq numdec} and \ref{eq nuadec} imply that
in this frequency range, for most realistic scenarios, it is a case
(i) light curve, i.e., $\nu_{a,dec},\nu_{m,dec}<\nuo$. Therefore,
Newtonian and mildly relativistic outflows as well as relativistic
GRB orphan afterglows peak at $t_{dec}$ with \citep{NP11}:
\begin{eqnarray}\label{eq Fpeak reg1}
   && F_{\nuo,peak}(\nu_{a,dec},\nu_{m,dec}<\nuo) \approx \nonumber\\
   &&0.3 {\rm ~m Jy~} E_{49} n^\frac{p+1}{4} \epsB^\frac{p+1}{4}
    \epse^{p-1} \bo^\frac{5p-7}{2} d_{27}^{-2} \left(\frac{\nuo}{1.4 {\rm
    ~GHz~}}\right)^{-\frac{p-1}{2}} .
\end{eqnarray}

The regime of $F_{\nuo,peak}$ at lower radio frequencies ($< 1$ GHz)
depends on the various parameters. If the outflow is Newtonian or
the density is low or the energy is low then
$\nu_{a,dec},\nu_{m,dec} < 100$ MHz and equation \ref{eq Fpeak reg1}
is applicable. Otherwise low radio frequencies are in regime (iii),
i.e., $\nuo<\nu_{eq},\nu_{a,dec}$. The flux peaks in this case at
\begin{eqnarray}\label{eq tpeak reg3}
   &&  t_{peak}(\nuo<\nu_{eq},\nu_{a,dec}) \approx\nonumber\\
   &&200 {\rm ~day~} E_{49}^\frac{5}{11}
    n^\frac{7}{22} \epsB^\frac{9}{22} \epse^\frac{6}{11} \left(\frac{\nuo}{150 {\rm
    ~MHz~}}\right)^{\frac{13}{11}} ,
\end{eqnarray}
with
\begin{eqnarray}\label{eq Fpeak reg3}
   &&  F_{\nuo,peak}(\nuo<\nu_{eq},\nu_{a,dec}) \approx\nonumber\\
   && 50 {\rm ~\mu Jy~} E_{49}^\frac{4}{5} n^\frac{1}{5} \epsB^\frac{1}{5}
    \epse^\frac{3}{5} d_{27}^{-2} \left(\frac{\nuo}{150 {\rm
    ~MHz~}}\right)^{\frac{6}{5}} .
\label{eq:fpeaklow}
\end{eqnarray}
In the last two equations we used  $p=2.5$ (other $p$ values in the
range 2.1-3 yield slightly different numerical factors and power
laws).

To date, the best observed signal from a mildly relativistic blast
wave is the radio emission that follows GRB associated SNe. The
main difference is that in these cases the circum burst medium is
typically a wind (i.e., $n \propto R^{-2}$) and therefore the
density at early times is much larger then in the ISM and self
absorption plays the main role in determining the light curve. A
good example for comparison of equation \ref{eq Fpeak reg1} with
observations is the light curve of SN 1998bw. This light curve is
observed at several frequencies at many epochs, enabling a detailed
modeling that results in tight constraints of the blast wave and
microphysical parameters. \cite{LiChevalier99} find that at the time
of the peak at $1.4$ GHz, about 40 days after the SN, taking
$\epsilon_e=\epsilon_B=0.1$, the energy in the blast wave is $\sim
10^{49}$ erg, its Lorentz factor is $\sim 2$ and the external
density at the shock radius is $n \sim 1 {\rm ~cm^{-3}}$. The peak
is observed when $\nu_m,\nu_a \leq \nuo$ and it depends only on
these parameters (it is only weakly sensitive to the density
profile). Therefore, equation \ref{eq Fpeak reg1} is applicable in
that case. Indeed, plugging these numbers into equation \ref{eq
Fpeak reg1} we obtain a flux of 20 mJy at the distance of SN 1998bw
(40 Mpc), compared to the observed flux of 30 mJy. This is not
surprising given that the model we use is based on that of radio
SNe.

\subsection{An outflow with a power-law velocity profile}\label{sec power-law profile}
The numerical simulations provide profiles of the outflow
energy as a function of the velocity. Similarly to the case of an
outflow with a single velocity we approximate the blast wave to be
spherical. The main difference when a range of velocities is
considered, is that in addition to the forward shock, which is
driven into the circum burst medium, there is a continuous reverse
shock that is driven into the ejecta. If most of the outflow energy
resides in low velocities then this reverse shock drives an
increasing amount of energy into the shocked region, mitigating the
deceleration of the forward shock. In that case the mass collected
by the forward shock, M(R), when its velocity is $\beta$ and its
radius is $R$, is comparable to the mass in the ejecta with velocity
$\geq \beta$. Thus the relation between the radius and the velocity
can be found at any time by equating:
\begin{equation}\label{eq Rbeta}
    M(R)(\be c)^2 = E(\geq \be) ,
\end{equation}
where this equality is correct up to a factor of order unity that
depends on the exact ejecta profile.
\cite{Chevalier82} calculates self-similar solutions for an outflow
with a power-law distribution $\rho \propto \be^{-k}$ in which most
of the energy is at low velocities, i.e., for $k >5$ since $E(>\be)
\propto \be^{-(k-5)}$. \cite{Chevalier82} provides the exact
coefficients of the solution for several values of $k$, all providing
corrections of order unity to equation \ref{eq Rbeta}.

In \S \ref{sec:lightcurves} we apply equation \ref{eq Rbeta} to the
results of the numerical simulations and we calculate the resulting
light curves from the ejecta profiles.  However before doing so, we
use equation \ref{eq Rbeta} to find $R(t)$ and $\beta(t)$ for an
outflow with a power-law velocity profile and we find an  analytic
solution for the flux for  a power-law velocity profile. Consider an
outflow with a power-law velocity profile and  a minimal velocity
$\be_{\min}$ and a total energy $E(>\be_{\min})=E_{tot}$, that
propagates in a constant density medium. Equation \ref{eq Rbeta}
implies that its radius and velocity evolve with time as:
\begin{eqnarray}\label{eq R selfsimilar}
    R &=&\left( \frac{3 k^{k-3} (\be_{\min} c)^{k-5} E_{tot}}{4 \pi (k-3)^{k-3} n m_p}
    \right)^{\frac{1}{k}} t^\frac{k-3}{k} \nonumber\\
    &\approx& 4 \cdot 10^{17}
    \left(\frac{E_{tot,50}}{n}\right)^\frac{1}{k}
    \left(\frac{\be_{\min}}{0.2}\right)^\frac{k-5}{k} t_{year}^\frac{k-3}{k}
    {\rm~cm~}
\end{eqnarray}
\begin{eqnarray}\label{eq beta selfsimilar}
    \be &=&\left( \frac{3 (k-3)^{3} (\be_{\min} c)^{k-5} E_{tot}}{4 \pi k^3 n m_p}
    \right)^{\frac{1}{k}} t^{-\frac{3}{k}}  \nonumber\\
    &\approx& 0.3 \left(\frac{E_{tot,50}}{n}\right)^\frac{1}{k}
    \left(\frac{\be_{\min}}{0.2}\right)^\frac{k-5}{k}
    t_{year}^{-\frac{3}{k}}
\end{eqnarray}
plugging these values into equations  \ref{eq num}-\ref{eq Fm} we
find that for relevant parameters $\nu_a, \nu_m < 1.4$ GHz and
\begin{eqnarray}\label{eq Fnu_Ev}
 \hspace*{-0.5cm}F_\nu(\nu>\nu_a,\nu_m) &\approx& 50 ~\mu{\rm Jy~} e^{-3.4(p-2.5)} E_{tot,50}^\frac{3+5p}{2k}
    \left(\frac{\be_{\min}}{0.2}\right)^\frac{(3+5p)(k-5)}{k} \nonumber \\
  &&n^\frac{5k-10p+pk-6}{k} \epsB^\frac{p+1}{4} \epse^{p-1} d_{27}^{-2}
    t_{year}^{\frac{3(2k-3-5p)}{2k}}\nonumber \\
    &=& 50 ~\mu{\rm Jy~} E_{tot,50}^{0.86}
    \left(\frac{\be_{\min}}{0.2}\right)^{3.44}
    n \epsB^{7/8} \epse^{3/2} d_{27}^{-2}
    t_{year}^{5/12}\nonumber \\
    && (k=9~;~p=2.5) .
\end{eqnarray}
This equation is applicable starting at $t_{dec}$ that corresponds
to the maximal velocity\footnote{We assume that the maximal velocity
ejecta is still mildly relativistic, i.e., $\g_{max} \be_{max}
\lesssim 1$.}, $\be_{max}$, and the energy that is carried by the
material that moves at the maximal velocity (i.e.,
$E_{\beta_{max}}\approx E_{tot}(\be_{max}/\be_{min})^{-k+5}$) and up
to $t_{dec}$ that corresponds to $\be_{min}$ and $E_{tot}$. At
earlier times then $t_{dec}(\be_{max},E_{\beta_{max}})$  the light
curve is the one described in table \ref{table1} (with $\beta_{max}$
and $E_{\beta_{max}}$) and at later times then
$t_{dec}(\be_{min},E_{tot})$ it joins the decaying light curve
described in table \ref{table1} ($F_\nu \propto t^{(21-15p)/10}$).

\subsection{Radio Light curves from the simulated outflows}
\label{sec:lightcurves} To calculate the electromagnetic signatures
we use the ejecta velocity profiles from the simulations and apply
the approximations of \S\ref{sec power-law profile}. We use equation
\ref{eq Rbeta} to find $R(t)$ and $\beta(t)$ and subsequently plug
them into equations \ref{eq num}-\ref{eq Fm} to calculate the light
curve. We approximate the ejecta-ambient medium interaction as a
spherical blast wave that propagates into a constant density, $n$.
Behind the shock constant fractions of the internal energy,
$\epsilon_e=0.1$ and $\epsilon_B=0.1$, are deposited in relativistic
electrons and in magnetic field, where the electrons are accelerated
to a power-law with an index $p=2.5$.

\begin{table*}
 \begin{minipage}{140mm}
  \caption{Basic properties of Radio Flares from the sub-relativistic dynamically ejected outflow for selected cases}
\begin{tabular}{|c||c|c|c|c|c|c|c|c|c|}
\hline
 Run  &  Masses &  \multicolumn{4}{|c|}{$n=1 $cm$^{-3}$}    & \multicolumn{4}{|c|}{$n=0.1 $cm$^{-3}$}          \\
&& \multicolumn{2}{|c|}{ 1.4 GHz}  &  \multicolumn{2}{|c|}{ 150 MHz}
&  \multicolumn{2}{|c|}{ 1.4 GHz}&
 \multicolumn{2}{|c|}{ 150 MHz}      \\
&&  $F_\nu$(peak$^a$)   &    t(peak$^a$)   & $F_\nu$(peak$^a$)  &    t(peak$^a$)   &$F_\nu$(peak$^a$)   &    t(peak$^a$)   &$F_\nu$(peak$^a$)&    t(peak$^a$)   \\
&$m_\odot$ &   mJy    & yr &    mJy    & yr &   $\mu$Jy    & yr &   $\mu$Jy    & yr    \\
\hline\hline
 8            & 1.4 -1.2& 0.09 & 4 & 0.5 & 4   & 10  & 9     & 50 & 9 \\
 12$^b$ &1.4 -1.4 & 0.04 & 1.5 & 0.2 & 2 & 5    & 3      & 30 & 3 \\
15$^c$  & 1.4 - 2.0    & 0.3 & 5 & 2 & 6    & 50  & 10    & 200 & 10 \\
23$^d$  &  1.4 - 10 & 1.5 & 4 & 4 & 8       & 200 & 10   & 1000 & 10 \\
\hline
\end{tabular}
\label{tab:radio_flares}
\newline $^a$ This is the peak of the sub-relativistic outflow. A mildly relativistic outflow, not calculated here,  may produce a stronger and earlier peak.
\newline $^b$ The canonical ns$^2$ case.
\newline $^c$ This is the maximal signal from our runs of ns$^2$ mergers
\newline $^d$ nsbh merger.
\end{minipage}
\end{table*}%

Inspection of the various ns$^2$ merger simulations reveal
relatively large kinetic energies (at least $10^{50}$ erg and at
times near $10^{51}$ erg, see Tab.~\ref{tab:cases}) but with
relatively low average velocities, around  0.1c - 0.16c. The
corresponding peak fluxes and durations correspond, therefore, to
the sub-relativistic case examined in \cite{NP11}. We calculate, for
these simulated mergers, the light curves at two frequencies, 1.4
GHz and 150 MHz, and for two values of external densities $n=1 {\rm
~ cm^{-3}}$ and $n=0.1 {\rm ~ cm^{-3}}$. The flux normalization is
for events at a distance of $10^{27}$ cm, roughly at the detection
horizon for ns$^2$ mergers by advanced LIGO and Virgo.

Here we calculate the radio emission only from the dynamical
component of the ejecta (see section \ref{sec:simulations}). This
component is launched preferentially along the equatorial plane,
while faster moving outflows (related to processes a  and c discussed
in section \ref{sec:simulations}) are launched along the rotation
axis. As a result the fast moving ejecta can propagate to large
distances, ahead of the dynamical ejecta, and interact with the
ambient medium. The faster moving components are expected to carry
less energy than the slow moving ejecta. Yet, the strong dependence
of the radio peak flux on the outflow velocity (equation \ref{eq
Fpeak reg1}) and the short deceleration time of the fast moving
ejecta (weeks to months) imply that  it is expected to dominate
radio emission before the dynamical ejecta start decelerating
(months-years). Thus, our estimates in this section   are only lower
limits {\em on} the true radio emission at early time (up to about a year
for $n=1 {\rm ~ cm^{-3}}$ and about three years for $n=0.1 {\rm ~
cm^{-3}}$).  A glimpse of the influence of a mildly relativistic
material can be seen in Figs. \ref{fig:light_curves_nsbh_n1} and
\ref{fig:light_curves_nsbh_n01} below which depicts the light curves
of nsbh mergers. Our simulations find that a mildly relativistic
material is ejected in the 1.4 $m_\odot$ ns - $5 m_\odot$ bh case. 
Although it carries a much lower energy
than the slower moving ejecta its strong effect is seen a early
times. Since we expect the dynamical ejecta to be the most massive
and most energetic of all the outflow components, it is not expected
to be affected significantly by possible interaction between the
different outflow components. If this is the case then our radio
predictions are good approximations to the true emission at later
times. If however, the fast outflow is more energetic than the slow
one, then also our late time radio estimates are only lower limits
of the true emission.

The resulting light curves are shown in Figs.
\ref{fig:light_curves_nsns_n1_eqm}-\ref{fig:light_curves_nsbh_n01}.
Figs. \ref{fig:light_curves_nsns_n1_eqm} and
\ref{fig:light_curves_nsns_n1 diffm} depict the radio light curves
of the various ns$^2$ mergers for $n=1~ {\rm cm}^{-3}$. Our
canonical case, a merger of two 1.4 $m_\odot$ neutron stars is
marked in both figures with a thick solid line.  The flux of the
canonical merger is 0.04mJy (0.2mJy) at 1.4GHz (150 MHz) 1-4 years
after the merger. It shows almost minimal emission among all equal
mass mergers (Fig. \ref{fig:light_curves_nsns_n1_eqm}), the emission
increases by a factor of 5 (at 1.4 GHz) for a 1.8-1.8 $m_\odot$ pair
and it is almost similar for 1.0-1.0 $m_\odot$. Fig.
\ref{fig:light_curves_nsns_n1 diffm} depicts mergers with
combinations of different ns masses. In mergers with a larger mass
difference the secondary is disrupted more completely, producing a
more prominent tidal tail. Therefore, the ejected mass rises with
the mass ratio (see Tab.~\ref{tab:cases} and Fig.
~\ref{fig:XYstructures}).  As a result, the dynamically ejected
outflows from unequal mass mergers, produce in general brighter and
longer lived radio remnants. Indeed, in most cases of unequal mass
mergers the luminosities, both in the 1.4 GHz and at 150 MHz, are
brighter at late times than the equal mass case. In some cases they
reach 0.3 mJy (2 mJy) a few years after the merger. Note that even a
small mass difference of less than $15\%$ can make a large
difference in the observed luminosity. For example, a 1.4-1.2
$m_\odot$ merger produces twice as bright and twice as long remnant
compared to a 1.4-1.4 $m_\odot$ merger.

The external density is a critical parameter. Figs.
\ref{fig:light_curves_nsns_n01_eqm} and
\ref{fig:light_curves_nsns_n01 diffm} depict the resulting light
curves when the density is $n=0.1~{\rm~ cm^{-3}}$ for $t>3$ yr. In
that case the signal, at the two considered frequencies, is almost
an order of magnitude fainter in all cases, compared with $n=1~{\rm
cm^{-3}}$. At $1.4$ GHz the signal  also evolves slower. Since the
self absorption frequency is below $150$ MHz in that case, the light
curves  at both 1.4 GHz and 150 MHz have  quite similar shapes.

Mergers of a black hole and a neutron star eject even larger amounts
of mass. Their signals, shown in Figs.
\ref{fig:light_curves_nsbh_n1} and \ref{fig:light_curves_nsbh_n01},
are stronger reaching 1 mJy at 1.4 GHz and a few mJy at 150 MHz. The
time scales are also longer and reach a few years at densities of 1
cm$^{-3}$ and about 10 year at 0.1 cm$^{-3}$.

\begin{figure}
\includegraphics[width=8.5cm,angle=0]{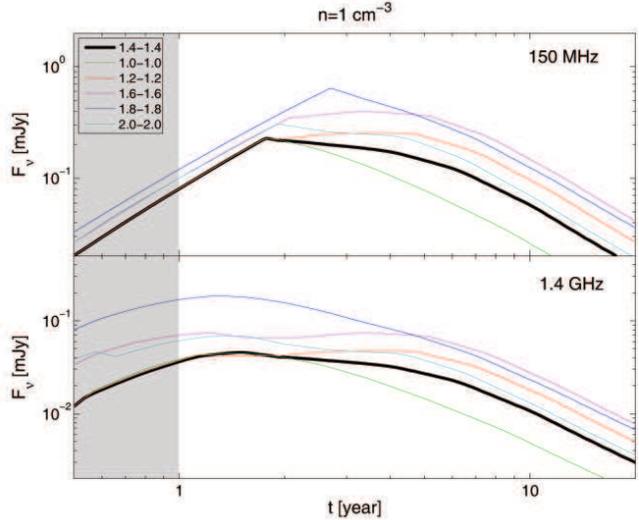}
\caption{Radio light curves generated by interaction of the
dynamically ejected sub-relativistic outflows from equal mass ns$^2$
mergers at 150 MHz and 1.4 GHz for 1 cm$^{-3}$ cirum-merger density,
$\epsilon_B=\epsilon_e=0.1$ and $p=2.5$. The merger distance is
$10^{27}$ cm, roughly the detection horizon for ns$^2$ mergers by
advanced LIGO and Virgo. The  shaded region at $t<1 $ yr reflect the
fact that at this time  the emission is expected to be dominated by
mildly relativistic outflows, which are not included in our
simulations.  }\label{fig:light_curves_nsns_n1_eqm}
\end{figure}

\begin{figure}
\includegraphics[width=8.5cm,angle=0]{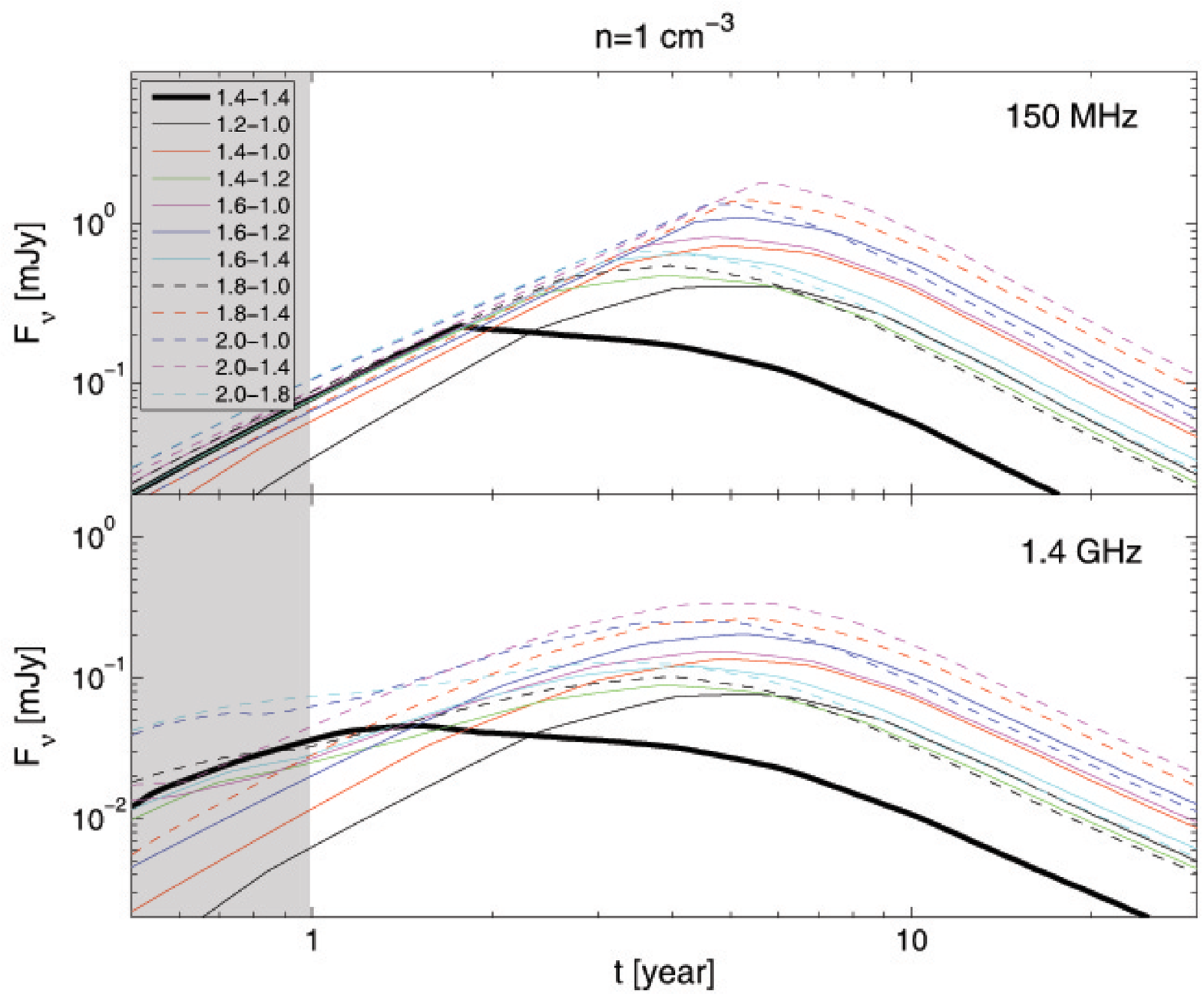}
\caption{Same as Fig. \ref{fig:light_curves_nsns_n1_eqm} for
different combinations of binary ns masses. } \label{fig:light_curves_nsns_n1 diffm}
\end{figure}

\begin{figure}
\includegraphics[width=8.5cm,angle=0]{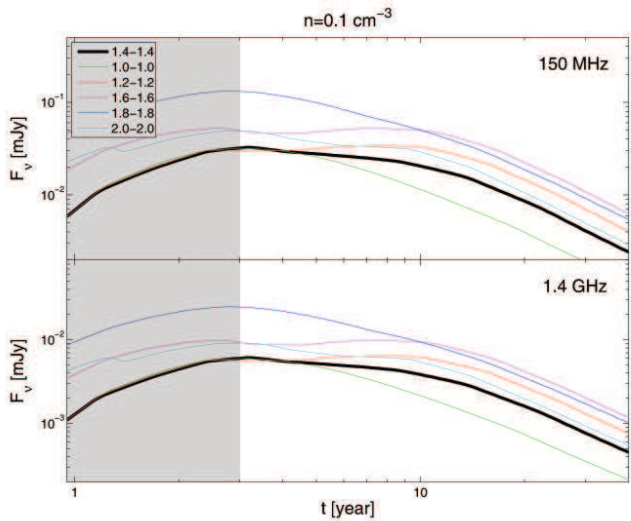}
\caption{Same as Fig. \ref{fig:light_curves_nsns_n1_eqm} for a equal
mass mergers  and density of  $n=0.1$cm$^{-3}$.
With the lower external density the effect of mildly relativistic
outflows extends now up to about 3 years and hence the shaded region
is larger. } \label{fig:light_curves_nsns_n01_eqm}
\end{figure}

\begin{figure}
\includegraphics[width=8.5cm,angle=0]{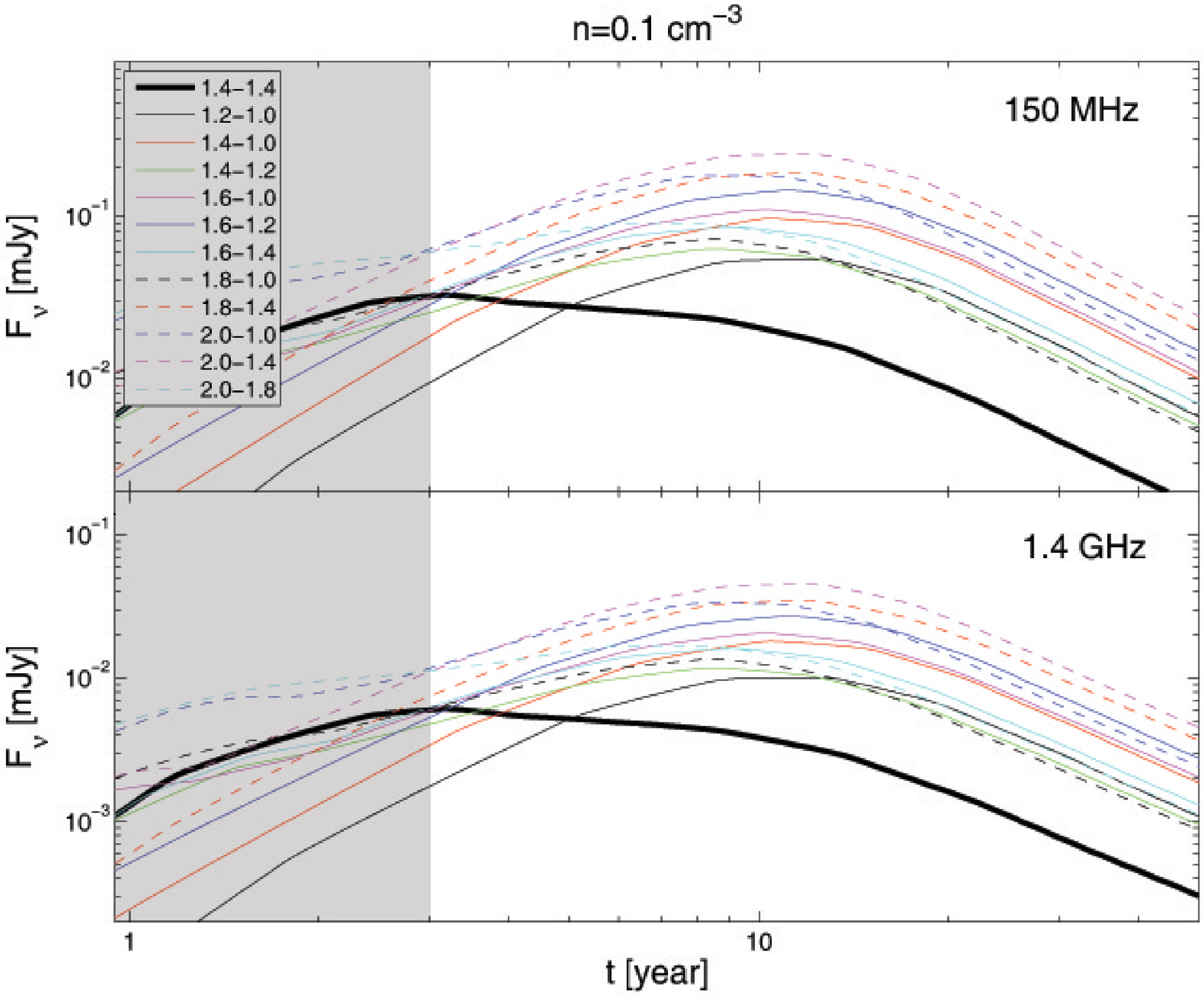}
\caption{Same as Fig. \ref{fig:light_curves_nsns_n01_eqm} for
different combinations of binary ns masses. }
\label{fig:light_curves_nsns_n01 diffm}
\end{figure}

\begin{figure}
\includegraphics[width=8.5cm,angle=0]{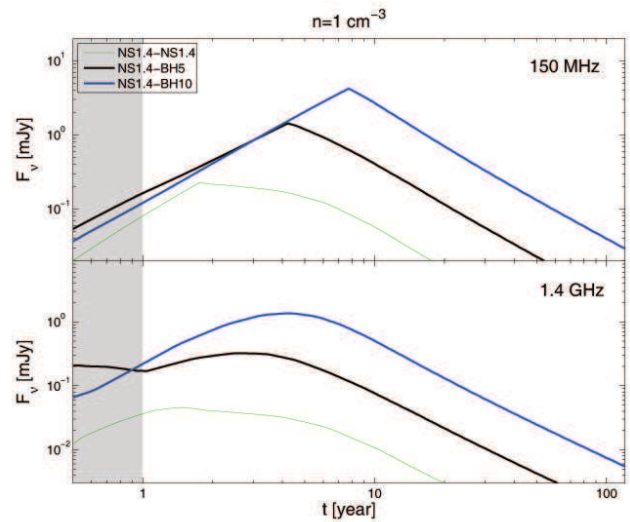}
\caption{Light curves for nsbh mergers of various masses. Parameters
are the same as Fig. \ref{fig:light_curves_nsns_n1_eqm}. For
comparison the result of the standard ns$^2$ merger case is shown as
a thin line.
 The strong effect of a modest mildly relativistic ejecta at early time can be seen 
in the 1.4$m_\odot$ ns-5 $m_\odot$ bh case in the 1.4 GHz lightcurve.
} \label{fig:light_curves_nsbh_n1}
\end{figure}

\begin{figure}
\includegraphics[width=8.5cm,angle=0]{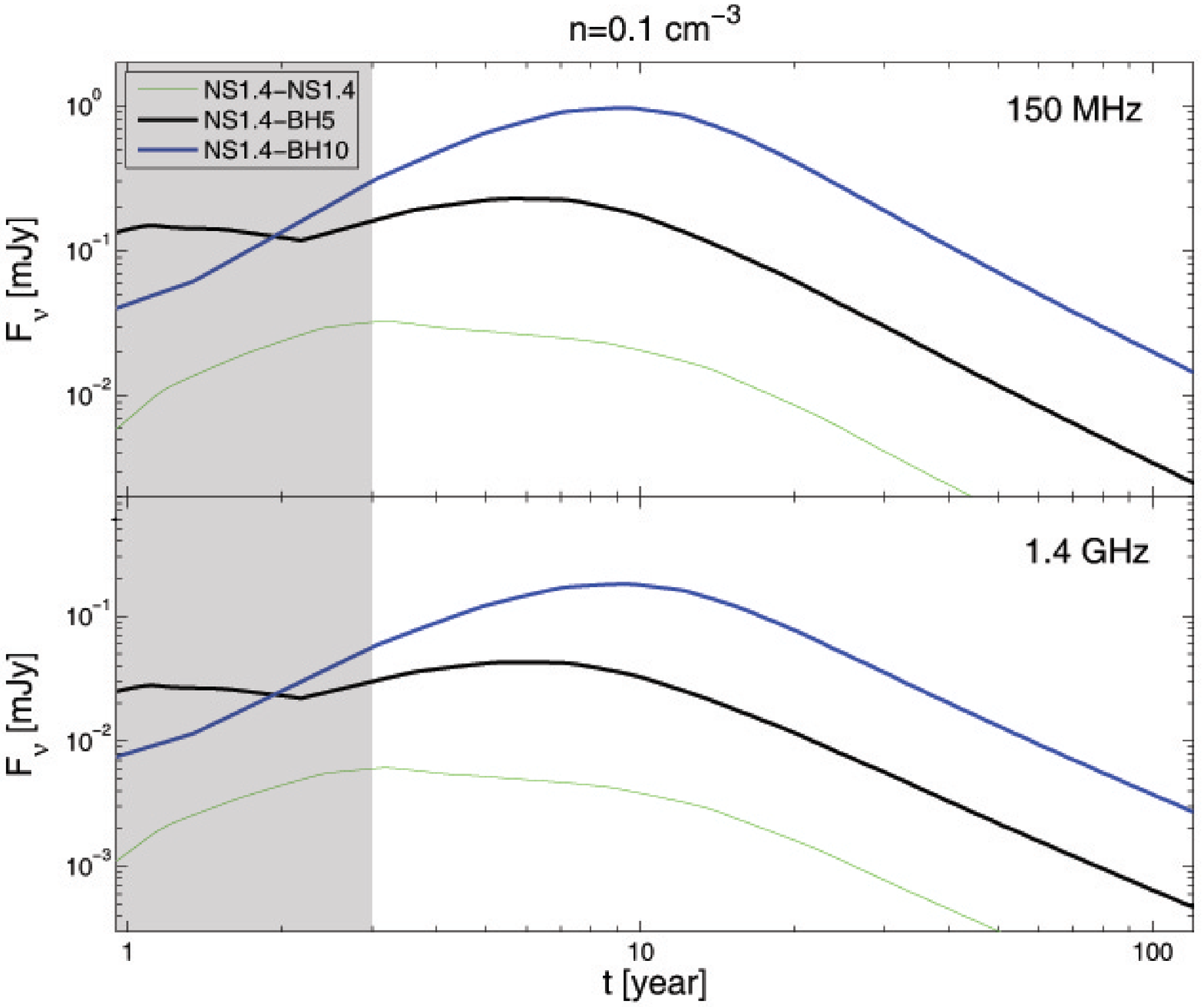}
\caption{Light curves for nsbh mergers of various masses. Parameters
are the same as Fig. \ref{fig:light_curves_nsns_n1_eqm} but the
density is 0.1 cm$^{-3}$.  The strong effect of a modest mildly relativistic ejecta at early time can be seen in the 1.4$m_\odot$ ns-5 $m_\odot$ bh case in both  150Mhz and 1.4 GHz lightcurves. } 
\label{fig:light_curves_nsbh_n01}
\end{figure}

\section{Detection and identification}
\label{sec:observations}
\subsection{Detectability}
We use the radio light curves of our canonical ns$^2$ merger
(1.4-1.4 \msun) to estimate their detectability by current and
future radio facilities. The estimates based on this merger case are
conservative.  First, since other merger cases, and especially those
with even a minor mass difference between the coalescing stars, are
brighter and second since our simulations do not include all outflow
sources. We also consider here the detectability of nsbh merger
based on the light curves calculated for a 1.4-10 \Msun merger.
Table \ref{tab:horizon} shows the detection horizons for these two
merger scenarios for different radio telescopes and two external
densities ($n=1$cm$^{-3}$ and $n=0.1$cm$^{-3}$).

The radio flares depend sensitively on the surrounding circum-merger
density. Current robust knowledge concerning ns$^2$ binaries arises
from the observed Galactic population. These binaries are all
observed in the Galactic disk whose typical density is about 1
cm$^{-3}$ \cite{Draine11}. Hence we expect that this is the most
relevant density to consider. In particular we stress that the
estimated rate of  ns$^2$, that are based on these binaries (and
used later when we estimate the detection rate of these radio
flares) are all relevant for this particular population. An
additional hypothetical population of ns$^2$ mergers that take place
in a much lower density environment and might produce weaker signals
is not included in our radio counterpart estimates.

A ns$^2$ merger that takes place in $n=1$cm$^{-3}$ environment
produces a radio remnant that can be easily observed by EVLA (1 hour
integration) all the way to the advanced LIGO/Virgo detection
horizon, with a rise time of $\sim$1 yr and a decay time of several
years. However, given the relative narrow EVLA field of view it
requires either a $\sim {\rm~deg^2}$ localization or a dedicated
search in a larger error box. Given that the planned localization of
advanced GW detectors is $\sim 10-100 {\rm~deg^2}$ for events with a
detector network signal to noise ratio of $10$ \citep{Wen10}, a
targeted search in the error box of a GW signal is certainly
feasible. For that purpose other facilities, with lower sensitivity
but significantly larger field-of-view (e.g., ASKAP) may be more
appropriate.

A ns$^2$ merger in $n=0.1$cm$^{-3}$ environment requires a more
dedicated effort (10 hours) in order to detect the resulting signal
out to $300$ Mpc by the EVLA. The rise and decay time are
significantly longer (years and a decade respectively). This makes
the search for ns$^2$ transients, even in the case of a $\sim 10-100
{\rm~deg^2}$ GW localization, very challenging, unless they produce
at least $10^{49}$ erg in mildly relativistic outflow (which is not
accounted for by our simulations).

nsbh mergers produce significantly brighter signals, and are
therefore detectable up to a greater distances, $\gtrsim 1$ Gpc, by
the EVLA and near future facilities, even for density of
$n=0.1$cm$^{-3}$. This distance is similar to the detection horizon
of these events by the advanced GW detectors. This makes nsbh
mergers a promising target, however, the observed rate depends of
course on their unknown merger rate.

Next we consider detectability in a blind radio survey. The
number of radio remnants in a single snapshot all sky radio image is
$N_{all-sky} ={\cal R} V \Delta t $, where $V$ is the detectable
volume, $\Delta t$ is the time that the flux is above the detection
limit and ${\cal R}$ is the event rate. Assuming a single velocity
outflow with typical parameters the $1.4$ GHz light curve is a case
(i) (see \S \ref{sec:Theory}). Thus, we use equations \ref{eq:tdec}
and \ref{eq Fpeak reg1}, and the approximation $\Delta t \approx
t_{dec}$, to find that the number of radio coalescence remnants in a
single $1.4$ GHz whole sky snapshot with a detection limit ${\cal
F}$ (see also supplementary material in \cite{NP11}:
\begin{equation}
\label{eq rate}
    N_{all-sky}^{1.4 {\rm GHz}} \approx 20 E_{49}^{11/6} n^\frac{9p+1}{24} \epsB^\frac{3(p+1)}{8}
    \epse^\frac{3(p-1)}{2}
    \bo^\frac{45p-83}{12} {\cal R}_{300}
    {\cal F}_{-1}^{-3/2} ~.
\end{equation}
Where   ${\cal F}_{-1}={\cal F}_{lim}/0.1$ mJy and ${\cal R}_{300}$
is the merger rate in units of $300 {\rm ~Gpc^{-3}~ yr^{-1}}$.
Plugging values from Table \ref{tab:cases} for the different runs we
can obtain a first estimate, including the dependence on the circum
merger density and microphysical parameters, of the number of all
sky flares that are detectable. A better estimate for specific
parameters can be obtained using the light curves we calculated in
\S \ref{sec:lightcurves}. Again we take 1.4-1.4 \Msun as the
canonical ns$^2$ merger and 1.4-10 \Msun as the canonical nsbh
merger.

A 1.4-1.4 \Msun ns$^2$ merger that takes place in $n=1$cm$^{-3}$
environment is brighter than $0.1$ mJy at 1.4 GHz for about 4 years
at a distance of 150 Mpc. Therefore, if as suggested by Galactic
ns$^2$, this is the density of a typical merger environment, then
$N_{all-sky}^{ns^2} \sim 20 {\cal R}^{ns^2}_{300} {\cal
F}_{-1}^{-3/2}$,
If however, typical ns$^2$ merger takes place at $n=0.1$cm$^{-3}$
environment the number of remnants in an all sky snapshot drops by
an order of magnitude.

\begin{table*}
 \centering
 \begin{minipage}{140mm}
\caption{ Properties and detection horizons (neglecting cosmological
corrections)  of the sub-relativistic dynamically ejected
outflow from 1.4-1.4 m$_\odot$ ns$^2$ and 1.4-10 m$_\odot$ nsbh
mergers with different radio facilities.}
\begin{center} {\footnotesize
\begin{tabular}{|c|c|c|c|c|c|c|c|}
\hline Radio   & Obs       & Field         & 1 hr         &  ns$^2$ 1 hr     &ns$^2$  10 hr  & nsbh 1 hr &   nsbh   10 hr   \\
       Facility    & Freq.     & of view      &  rms        &   horizon$^\dag$ & horizon$^{\dag\dag}$ &   horizon$^\dag$ &   horizon$^{\dag\dag}$  \\
                      &  (GHz)   &    (deg$^2$)    & $\mu$Jy  &$n=1$cm$^{-3}$  & $n=0.1$cm$^{-3}$ &$n=1$cm$^{-3}$  & $n=0.1$cm$^{-3}$ \\
\hline\hline
EVLA$^a$ & 1.4  & 0.25      & 7     &         360 Mpc  & 200Mpc &1.8 Gpc & 1.4 Gpc   \\
ASKAP$^b$& 1.4  & 30    & 30    &    170 Mpc    &  100 Mpc & 850Mpc& 700 Mpc\\
MeerKAT$^c$& 1.4  & 1.5     & 35    &    160 Mpc  &  90 Mpc & 800 Mpc & 650 Mpc\\
Apertif$^d$ &1.4  & 8       & 50    &   135 Mpc &   75 Mpc& 670 Mpc & 550 Mpc\\
LOFAR$^e$ & 0.15 & 20   &  1000 & 70 Mpc & 40 Mpc & 300Mpc & 250 Mpc\\
\hline
\end{tabular} }
\end{center}
$^\dag$ The distance at which the observed peak flux is 4 times the
1 hr rms.
\newline$^{\dag\dag}$ The distance at which the observed peak flux is 4 times the
10 hr rms.
\newline$^a$ http://www.aoc.nrao.edu/evla/
\newline $^b$ http://www.astron.nl/general/apertif/apertif
\newline $^c$ http://www.atnf.csiro.au/projects/askap/technology.html
\newline $^d$ http://www.ska.ac.za/meerkat/
\newline $^e$ http://lofar.org
\label{tab:horizon}
\end{minipage}
\end{table*}%

A 1.4-10 \Msun  ~nsbh merger that takes place in $n=1$cm$^{-3}$
environment is brighter than $0.1$ mJy at 1.4 GHz for about 4 years
at a distance of 1 Gpc, implying $N_{all-sky}^{nsbh} \sim 20 {\cal
R}^{nsbh} {\cal F}_{-1}^{-3/2}$, where ${\cal R}^{nsbh}$ is the nsbh
merger rate in units of ${\rm ~Gpc^{-3}~ yr^{-1}}$. Again, if the
typical nsbh circum-merger density is $n=0.1$cm$^{-3}$ then
$N_{all-sky}$ drops by an order of magnitude.

To conclude, our calculations here strengthen the results of
\cite{NP11} that  there  is a fair chance to detect merger radio
remnants in a sub-mJy survey of even a part of the sky and a high
chance to detect them in a whole sky survey. Such survey's of a
small portion of the sky are planed already with the EVLA, while a
sub-mJy large scale transient survey is part of the ASKAP Survey
Science
Projects\footnote{http://www.atnf.csiro.au/projects/askap/ssps.html}.
Finally, a 1mJy 150 MHz survey with LOFAR will find a comparable
number of remnants, but these will have a longer rise time, and
therefore will be harder to identify.

\subsection{A comparison with short GRBs' radio orphan afterglows}
The outflows of short GRBs begin highly relativistic and probably highly
beamed. Eventually they slow down and become detectable from all
directions (see \S \ref{sec:Theory}). Therefore, the rate estimate
equation  \ref{eq rate} is also applicable for radio orphan afterglows
when $\beta_0 = 1$. However some of the parameters in equation
\ref{eq rate} are not directly observable. The observed quantities
are the isotropic equivalent $\g$-ray energy, $E_{\g,iso}$, and the
rate of bursts that point to the observer ${\cal R}^{SHB}_{obs}$,
while equation \ref{eq rate} depends on $E = E_{iso} f_b$ and
${\cal R^{SHB}} = {\cal R}_{obs}^{SHB} f_b^{-1}$, where $f_b<1$ is
the fraction of the $4 \pi$ steradian covered by the jet and $E_{iso}$
is the isotropic equivalent energy in the afterglow blast wave.
X-ray observations indicate that $\g$-ray emission in short GRBs is
very efficient and that in general $E_{iso} \sim E_{\g,iso}$ \citep{Nakar07}.
In the following discussion we assume that this is the case.

$E_{\g,iso}$ of short GRBs ranges at least over four orders of
magnitude ($10^{49}-10^{53}$ erg). The rate of observed short GRBs
is dominated by $10^{49}$ erg bursts, and the luminosity function
can be well approximated by a power-law, at least in the range $\sim
10^{49}-10^{51}$ erg, such that ${\cal R}_{obs}^{SHB}(E) \sim 10
E_{iso,49}^{-\alpha} {\rm ~Gpc^{-3}~yr^{-1}}$ where $\alpha \approx
0.5-1$ \citep{NGF06,GP06}. Plugging these into equation \ref{eq
rate} we obtain
\begin{equation}\label{eq orphan_rate}
    N_{all-sky~SHB}^{1.4 {\rm GHz}} \approx 1 f_b^{5/6}
    E_{iso,49}^{\frac{11}{6}-\alpha} n^\frac{9p+1}{24} \epsB^\frac{3(p+1)}{8}
    \epse^\frac{3(p-1)}{2}  F_{lim,-1}^{-3/2} ~.
    \label{eq:orphan}
\end{equation}
This equation is similar to equation 9 of \cite{Levinson+02}, with
the observed luminosity function already folded in.

Narrower beamed bursts (with lower $f_b$) are more numerous and they
produce less total energy per burst.  The positive dependence of
equation \ref{eq orphan_rate} on $f_b$ implies that overall the
lower energy is ``winning'' over the increased rate, and the
detectability of narrower bursts is lower. Using, equation \ref{eq
orphan_rate} we can put a robust upper-limit on the orphans rate
since all the parameters are rather well constrained by
observations, with the exception of $f_b$ which is $<1$ by
definition. Therefore, assuming that short GRBs are beamed, the
detection of the common $\sim 10^{49}$ erg bursts in a blind survey,
even with next generation radio facilities, is unlikely
\citep{Nakar07}. However, brighter events should be detectable. If
the beaming is energy independent, detectability increases with the
burst energy.  The luminosity function possibly breaks around
$10^{51}$ erg, in which case the orphans number is dominated by
$10^{51}$ erg bursts. For $f_b^{-1}=30$ we expect from these bursts
$\sim 10$ orphan afterglows brighter than 0.1 mJy at a single $1.4$
GHz whole sky snapshot.

So far we discussed detectability in a blind survey. A followup
dedicated search would be, of course, more sensitive. If compact
binary mergers produce short GRBs then the energy of most GW
detected bursts will be faint with $E_{\g,iso} \sim 10^{49}$ erg.
The chance to detect their orphan afterglows again depended on their
total energy and thus on $f_b$. Equation \ref{eq Fpeak reg1} shows
that if $f_b^{-1}=30$ then detection should be difficult but
possible in a dedicated search mode. Note that since the energy of
the burst is low, the radio emission will evolve quickly, reaching a
peak and decaying on a week time scale, so a prompt and rather deep
search will be needed.

\subsection{Identification and contamination}

A key issue with the detection of compact binary merger remnants in
blind surveys is their identification.  \cite{Ofek+11} and
\cite{Frail+12}  present a census of the transient radio sky.
Luckily the transient radio sky at $1.4$ GHz is relatively quiet.
The main contamination source are radio active Galactic nuclei
(AGNs), however their persistent emission is typically detectable in
other wavelength and/or deeper radio observations. Moreover, the
signal from a compact binary merger is expected to be located within
its host galaxy (otherwise the density is too low), but away from
its center. The host and the burst location within it, should be
easily detectable at the relevant distances.

The only known, and guaranteed, transient 1.4 GHz source with
similar properties are radio SNe. Among these typical radio SNe are
the most abundant. A transient search over 1/17 of the sky with
$F_{lim}=6$ mJy at 1.4GHz \citep{Levinson+02,Gal-Yam+06,Ofek+10}
finds one radio SN. This rate translates to $10^{3}-10^{4}$ SNe in a
whole sky $F_{lim}=0.1$ mJy survey. These contaminators can be
filtered in three ways. First, by detection of the SN optical light.
However, the optical signal may be missed if it is heavily
extincted, and given the large number of radio SNe, misidentifying
even a small fraction of them may render the survey useless for our
purpose. The second filter is the optically thick spectrum at high
radio frequency ($\sim 10$ GHz) at early times, which is a result of
the blast wave propagation in a wind. Thus, a multi-wavelength radio
survey can identify radio SNe. The last filter is the
luminosity-time scale relation of typical radio SNe that is induced
by the outflow velocity \citep[e.g., Fig. 2 in][]{Chevalier+06b}.
Type II SN outflows are slow, $\sim 0.01$c, and therefore their
radio emission is longer/fainter than that expected for merger
remnants. The common type of Ib/c radio SNe is produced by $\sim
0.2$c blast waves but with much less energy than what we expect from
a binary merger outflow, and therefore their radio emission is much
fainter. The combination of any two of these filters will hopefully
be enough to identify all the typical radio SNe.

Slightly different contaminators are GRBs associated SNe. Their
outflows are as fast and as energetic as those that we expect from a
binary merger and therefore their radio signature is similar in time
scales and luminosities. SN1998bw-like events are detectable by a
0.1 mJy survey at 1.4 GHz up to a distance of several hundred Mpc
for 40 days and their rate is $40-700 {\rm ~Gpc^{-3} ~yr^{-1}}$
\citep{skn+06}, implying at least several sources at any whole sky
snapshot. Here only the first filter (SN optical light) and possibly
the second (optically thick spectrum) can be applied. However, given
the high optical luminosity of GRBs associated SNe and their
relatively low number this should be enough in order to filter them
out. These contaminators highlight the importance of a multi-wave
length strategy where an optical survey accompanies the radio survey
to best utilize both surveys' detections.

Finally, radio is the place to look for blast waves in tenuous
mediums, regardless of their origin. Any source of such an
explosion, be it a binary merger, a GRB or a SN, produces a radio
signature. Therefore, all the strong explosions may be detectable in
a deep radio survey, this includes for example long GRB on-axis and
off-axis afterglows and giant flares from extragalactic soft
gamma-repeaters. The difference between the radio signatures of the
different sources (amplitude, spectrum and time evolution) depends
on the blast wave energy and velocity and on the external medium
properties. We thus will be able to identify the characteristics of
binary mergers outflows. If, however, there is a different source of
$\sim 10^{50}$ erg of sub- or mildly relativistic outflow that
explodes in the ISM it will be indistinguishable from binary mergers
(at least in the radio). Currently we are not aware of any such
source, with the exception of long GRBs at the low end of the
luminosity function, but these are too rare to contaminate a survey.
Any other source of such outflows, if existent, will probably be a
part of the family of collapsing/coalescing compact objects.

\section{IR-UV transients from radioactive decays, ``macronovae''}
\label{sec:macronova}

The ejected material is extremely neutron rich
and rapidly expanding. Under such conditions rapid neutron capture
is hard to avoid \citep{hoffman96,freiburghaus99b,roberts11,korobkin12}.
The latter study finds that ns$^2$ and nsbh mergers produce a unique,
   solar-system-like r-process abundance pattern for nucleon numbers
   $A > 120$, independent of the astrophysical parameters of the merging
   binary system.  The
r-process itself occurs on a very short time scale, but the freshly
synthesized elements subsequently undergo nuclear fission, alpha-
and beta-decay which occur on much longer time scales. The
supernova-like emission from this expanding material was first
suggested by \cite{LP98} and discussed later  by \cite{Kulkarni05}
and \cite{metzger10b}. We combine the ejecta velocity profiles found
in our simulations with the time dependent radioactive power
injection found in the recent r-process study of \cite{korobkin12}
to calculate bolometric light curves.

The optical depth of the expanding outflow decreases with time, and
as a result larger amounts of mass become visible to the observer.
The ejected mass has a gradient of velocities. We denote as $m(v)$
the mass with asymptotic velocity at infinity $>v$. Before the peak of the
emission, the observed mass, $m_{obs}$, is the one for which the
diffusion time is comparable to the expansion time, namely
$\tau_{obs} \approx c/v$, where $\tau(m) \approx \kappa m/(4\pi v^2
t^2)$ (note that $m=m(v)$) and $\kappa$ is the opacity cross-section
per unit of mass. The emission peaks at $t_{peak}$, when the entire
ejecta become observable, namely at the first time that
$m_{obs}=m_{ej}$. At later time the entire ejecta are exposed and
$m_{obs}(t>t_{peak})=m_{ej}$.

A major uncertainty in the light curve calculation is the
opacity of r-process elements. The opacity determines first the
bolometric luminosity and second its spectrum. Here we assume a
constant and grey opacity in order to determine the luminosity,
which is less sensitive to the detailed wavelength dependence of the
opacity. We do not attempt to predict the observed spectrum.}
\cite{metzger10b} discussed the opacity of the neutron rich ejecta
expect it to be similar to that of iron-group elements, which they
approximate as a constant $\kappa= 0.1$ cm$^2$/g. Below use it as
the canonical value.

At any time prior to $t_{peak}$ the observed mass is
\begin{equation}\label{eq:mobs}
   m_{obs}(t<t_{peak})\approx 0.05 M_\odot \left(\frac{\kappa}{0.1{\rm cm^2/g}}
   \right)^{-1}\left(\frac{t}{{\rm day}} \right)^2 \left(\frac{v}{0.1 c}\right)  .
\end{equation}
Note that this is an implicit equation since $v$ itself depends on
$m_{obs}$. Thus, $m_{obs}$ increases with time (slightly slower than
$t^2$, since $v$ decreases with $m$) until all of the ejecta is
exposed at $t_{peak}$. The observed luminosity is dominated by the
energy release via radioactive decay. \cite{korobkin12} follow the
nucleosynthesis of a number of fluid elements that are ejected at
different velocities and during different stages of the merger and
find that all of them result in a similar radioactive energy
injection rate per unit of mass into the expanding ejecta that can
be approximated by
\begin{equation}\label{eq edot}
\dot{e}=2 \cdot 10^{18} \left[\frac{1}{2}-\frac{1}{\pi}\arctan\left(
\frac{t-t_0}{\sigma}  \right) \right]^{1.3} \times
\frac{\epsilon_{therm}}{0.5} \rm{~ erg/g/s}
\end{equation}
where $t_0=1.3$ s and $\sigma=0.11$ s and $\epsilon_{therm}$ is the
fraction of injected energy that is emitted as thermalized radiation
and not as unobserved neutrinos or gamma-rays. The value of
$\epsilon_{therm}$ is not well constrained and below we use the default value of \cite{korobkin12},
 $\epsilon_{\rm therm}= 0.5$. On the relevant
time scales equation \ref{eq edot} corresponds to the \cite{LP98}
coefficient $f= 10^{-6} (t/{\rm day})^{-0.3}$. The resulting
bolometric (thermalized) luminosity is:
\begin{equation}\label{eq:L(t)}
   L \sim m\dot{e} \sim 2 \times 10^{41}
   {\rm~erg~s^{-1}}\frac{\epsilon_{therm}}{0.5} \left(\frac{m_{obs}}{10^{-2}M_\odot}\right) \left(\frac{t}{\rm day}
   \right)^{-1.3}.
\end{equation}
Note that at early times $L$ increases because $m_{obs}$  increases
almost like $t^2$.  This luminosity is emitted in the  IR to UV range.
At early time $L$ is increasing (slightly slower than linearly). The
luminosity peaks when the entire ejecta is seen (roughly at $t=0.5$
day), and the light curve decays afterwards roughly as $t^{-1.3}$.
The value of $m_{obs}$ is inversely proportional to the
opacity. Therefore higher opacity implies a slower evolving and
dimmer signal, and it most likely results in a redder transient.

\begin{table}
 \begin{minipage}{140mm}
  \caption{Basic properties of Macronovae for selected cases}
\begin{tabular}{|c||c|c|c|c|c|c|c|c|c|}
\hline
 Run  &  Masses    &  $m_{\rm ej}$  & L(peak)$^\dag$           &    t(peak)   & magnitude$^\dag$    \\
            &$m_\odot$ &   $10^{-2}$\msun            & $10^{41}$ erg/s    & day            &     at 300 Mpc                          \\
\hline\hline
 8            & 1.4 - 1.2   & $2.1 $ & 6 [1.5]    & 0.4 [1.5] & 22 [24] \\
 12$^a$ & 1.4 - 1.4   & $1.3 $ & 5 [1] & 0.4 [1.5]& 22 [24]     \\
15$^b$         & 1.4 - 2.0  & $3.9 $ & 9 [2] & 0.6 [2] & 21.5 [23.5]  \\
23$^c$  & 1.4 - 10    & $4.9 $ & 11 [2.5] & 0.6 [2] & 21 [23]      \\
\hline
\end{tabular}
\label{tab:Macronovae}
\newline All values are assuming   $\kappa=0.1 {\rm
~cm^2/g}$ [$\kappa=1 {\rm ~cm^2/g}$]
\newline $^\dag$  Bolometric thermalized luminosity/magnitude, emitted mostly in
IR, optical and UV.
\newline $^a$ The canonical ns$^2$ case.
\newline $^b$ This is the maximal signal among the ns$^2$ mergers runs.
\newline $^c$ nsbh merger.
\end{minipage}
\end{table}%

\begin{figure}
   \includegraphics[width=8.5cm,angle=0]{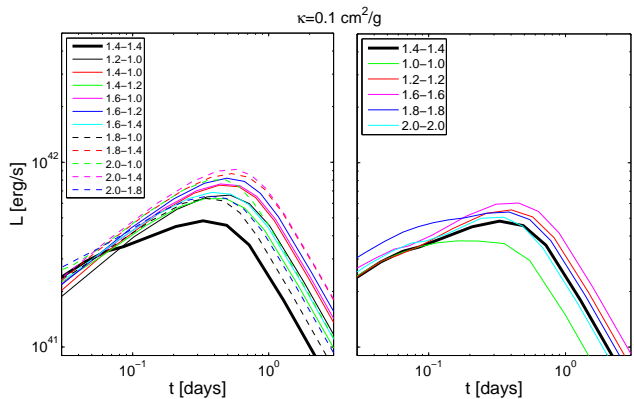}
   \caption{Bolometric light curves of  "macronovae"  - supernova-like
            event that are powered by a radioactive decays within the
            dynamical ejecta. The opacity is assumed to be grey and constant with $\kappa=0.1 {\rm
~cm^2/g}$. Most of the luminosity is emitted in IR, Optical and UV.
            Left panel: different ns masses; Right panel:
            equal ns masses. The bold solid black line corresponds to the
            canonical case of a 1.4-1.4 \Msun merger.}
   \label{fig:macronova_lightcurves}
\end{figure}

\begin{figure}
   \includegraphics[width=8.5cm,angle=0]{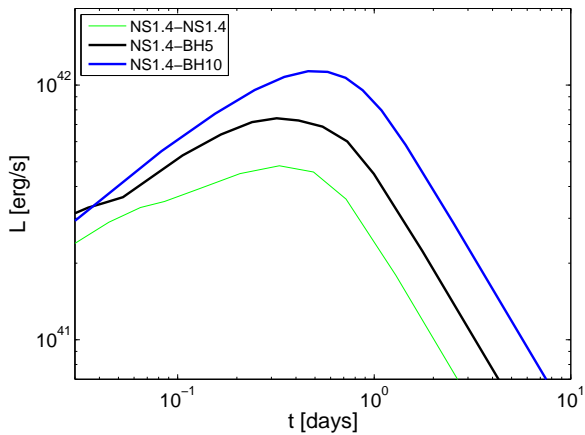}
   \caption{Bolometric light curves of  "macronovae"  - supernova-like
            event that are powered by a radioactive decays within the
            dynamical ejecta. The opacity is assumed to be grey and constant with $\kappa=0.1 {\rm
~cm^2/g}$. Most of the luminosity is emitted in IR, Optical and UV.
            The lightcurves are for nsbh mergers of
            different masses. For a comparison the light curve for the
            canonical 1.4-1.4 \Msun merger is also shown.}
   \label{fig:nsbh_macronova_lightcurves}
\end{figure}

\begin{figure}
   \includegraphics[width=9cm,angle=0]{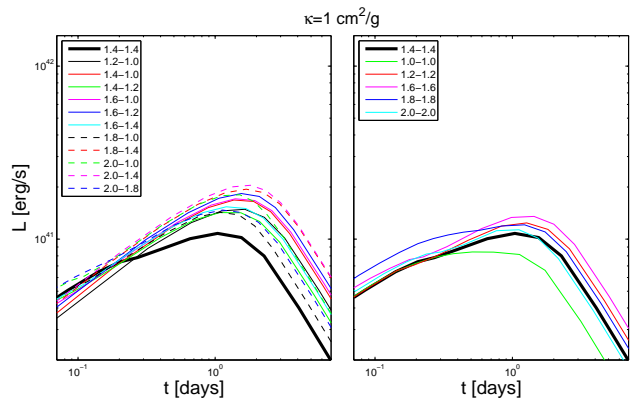}
   \caption{Same as figure \ref{fig:macronova_lightcurves} with $\kappa=1 {\rm
~cm^2/g}$. The resulting light curves are dimmer and evolving more
slowly for larger values of $\kappa$.}
   \label{fig:macronova_lightcurves_k=1}
\end{figure}

The resulting macronova light curves, assuming   $\kappa=0.1 {\rm
~cm^2/g}$, are shown in Fig.~\ref{fig:macronova_lightcurves} for
different  ns$^2$  mergers. The canonical merger case (thick black
line) peaks after 0.4 days with $5 \times 10^{41}$ erg/s.  The
macronovae light curves produced by the two nsbh mergers that we
consider are depicted in Fig.~\ref{fig:nsbh_macronova_lightcurves},
the canonical ns$^2$ case (2 $\times 1.4$ \msun) is shown for
reference purposes. Since more mass is ejected these macronovae peak
later, at about 0.7 days and their peak luminosities can reach about
$10^{42}$ erg/sec. Our estimates of the macronova peak luminosity
for the canonical ns$^2$ is similar to the calculation of
\cite{Metzger+10} for $m_{ej}=0.01$ \Msun and $v=0.1c$. Other merger
cases are brighter by a factor of 2-3 mostly due to the larger
amount of ejected mass.  Figure \ref{fig:macronova_lightcurves_k=1} show the effect of the opacity on the light curves. Increased
opacity delay the peak time ($\propto \kappa^{0.5}$) and reduce its
luminosity ($\propto \kappa^{-0.65}$).

A peak luminosity of $\sim 5 \times 10^{41}$ [$ 10^{42}$] erg, that
arises in our canonical ns$^2$ [10 \Msun nsbh] merger, corresponds
to  an absolute bolometric magnitude of about -15 [-16], or an
observed bolometric magnitude of 22 [21] at a distance of $\sim
300$Mpc. The observed magnitude at a given observed optical band,
which does not necessarily include the peak of the signal spectrum
will be fainter.
Thus, a more realistic estimate of the observed luminosity at a
distance of $\sim 300$ Mpc is 22.5-23.5[21.5-22.5]. The nsbh
macronova bolometric magnitude would be 23.5 at 1 Gpc, roughly the
LIGO/Virgo detection horizon for the nsbh merger's GW signal and at
a given optical band it will probably be a magnitude or two fainter.
With a rapid follow-up this can be easily picked up within the
duration of 0.4 days by various optical telescopes. However, the
large field of view needed to be covered within a short period of
time (of order of 100 deg$^2$) would be very challenging (see
discussion by \citealt{MetzgerBerger12}). Detection by a blind
survey is also possible. Using the PTF limiting magnitude of 21
\citep{Law09,Rau09} a ns$^2$ macronova can be detected by the PTF to
a distance of $\sim 100$ Mpc. The slightly stronger nsbh macronova
could be detected to a distance of $\sim 200$ Mpc. With a cadence of
5 days  only one in twelve ns$^2$ macronova, that last 0.4 days,
will be detected. Taking a covered region of $\sim 2000$ deg$^2$ the
expected detection rate is $0.004 {\cal R}_{300}$ macronova per
year. Detection by the LSST is much more likely. With a full sky
coverage at 24.5 magnitude and a cadence of 3 days we expect a LSST
horizon for a ns$^2$ macronova at about 500 Mpc and a detection rate
of $\sim 20 {\cal R}_{300}$ ns$^2$ macronovae per year. However,
most of those will be observed as single detection transients and
identifying their nature is a nontrivial challenge.  \\
\\

\section{Conclusion}
\label{sec:conclusions}

Compact binary mergers are expected to eject sub-relativistic, mildly
relativistic and possibly ultra-relativistic outflows. The interaction of such an  outflow with the
circum-merger environment produces a long-lived radio remnant.
Radioactive decay within the outflow produces a short lived IR-UV
transient.

To estimate the properties of these transients we have carried out a
series of merger simulations which aims at finding the properties of
the  dynamically ejected mass. This is a lower limit on the amount
of the ejected mass, since other sources of sub-, mildly and 
ultra-relativistic outflows are not accounted for in our simulations,
see the discussion at the end of Sec.~\ref{sec:simulations}. 
We find that our canonical merger case of two 1.4 $m_\odot$ neutron stars
dynamically ejects 0.013 \Msun with a distribution of
velocities in the range of 0.05-0.2 c and an average value $\langle
v \rangle \approx 0.1$c.  The energy carried by this outflow is $1.6
\times 10^{50}$ erg. Other ns$^2$ mergers, and especially those with
unequal masses, generate more massive outflows at slightly faster
velocities, up to 0.04 \Msun and $9 \times 10^{50}$ erg of kinetic
energy in some cases. The dynamically ejected outflow that we find
for nsbh mergers carries about $10^{51}$ erg.

A strong radio  remnant is expected in any merger of a ns$^2$ whose
properties are similar to those of the Galactic neutron star binary
population (i.e., typical circum-merger Galactic disk density of $1
{\rm~cm^{-3}}$). Such a radio remnant appears months to years after
the merger and remains bright for a similar time. Therefore, a
trigger following the detection of a GW signal can wait weeks after
the event and no online triggering is needed. In addition the long
lifetime of the  remnants enables their detection in a blind survey.
Here we calculate the light curves from one year after the
merger, since our simulations do not include the mildly relativistic
outflow which dominates the emission at earlier times. The radio
flux from a canonical 1.4-1.4 $m_\odot$ ns$^2$ merger taking place
at the advanced LIGO/Virgo horizon for such mergers, $300 $ Mpc,  in
$n=1$ cm$^{-3}$ environment is 0.04 mJy (0.2 mJy) at 1.4 GHz (150
MHz) 1-4 years after the merger. This signal could be easily
detected by an one hour observation of the EVLA or by a whole day
observation on ASKAP or MeerKAT. The sensitivity of present lower
frequency detectors, e.g. LOFAR, at  150 MHz, is insufficient for a
detection at the advanced LIGO/Virgo horizon. Longer observations
(e.g. 10 hours on the EVLA) can detect these mergers even if they
are in a lower density environment ($n=0.1$ cm$^{-3}$). A mildly
relativistic component in the ejecta probably increases the
brightness and detectability of the signal on time scales of
weeks-months. The nsbh GW horizon is farther than the ns$^2$ GW
horizon. Our numerical simulations find that nsbh mergers produce
higher energy outflows resulting in larger fluxes. Overall we find
that the ELVA detection horizon for nsbh is almost a factor of two
larger than the advanced LIGO/Virgo detection horizon.

We find that the optimal frequency to carry out a search for
merger remnants is $1.4$ GHz. Taking a sub-relativistic outflow
with $10^{50}$ erg  and a canonical ns$^2$
merger rate of 300 Gpc$^{-3}$ yr$^{-1}$(and a range of 20 - $2
\times 10^4$ Gpc$^{-3}$ yr$^{-1}$) we expect a detection of $\sim 20$
(1-1200 correspondingly) radio ns$^2$ remnants in a $0.1$ mJy all
sky survey. The expected higher velocity component
increases this rate making remnants detectable even in a survey
that covers only a small fraction of the sky or that operates at a mJy
sensitivity. Thus, a sensitive large field-of-view GHz survey by
currently available facilities has a great potential to constrain
the rate of binary mergers, information that is of great importance
for the design and operation of the advanced  GW detectors.

The detectability of the radio transients depends strongly on the
circum-merger density, which may be low if the binary has been
ejected from its host galaxy before the merger. This uncertainty
implies that there may be a fraction of ns$^2$ mergers, that take
place out of the disk of Milky Way-like galaxies,  whose     radio
flares are faint. However our estimates of the detection rate in a
blind radio survey are less  affected by this uncertainty as our
canonical circum-merger density and the expected merger rate are
based on the observed Galactic ns$^2$ binaries, which are all
located in the Galactic disk. If there is a population of mergers
which take place out of the disk of Milky Way-like galaxies, it will
be in addition to the population that we consider here for radio
transient blind searches.

These radio transients should be compared with  short GRB orphan
radio afterglows. These may be produced by compact binary
mergers if they are launching also ultra-relativistic outflows.
Our estimates of orphan afterglows detectability are based on short
GRB observations and are therefore independent of whether short GRBs
are binary mergers or not. The main uncertainty in the rate
estimates is the GRB beaming factor. We find that assuming
$f_b^{-1}=30$ there are about a dozen orphan afterglows at 0.1 mJy
in a single $1.4$ GHz whole sky snapshot. These are dominated by
relatively energetic short GRBs ($E_{\g,iso} \sim 10^{51}$) and
their duration is several weeks. If binary mergers are short GRB
engines then a GW-triggered event will most likely be of a low
energy GRBs, $E_{\g,iso} \sim 10^{49}$ erg, and a true energy, after
beaming correction, that is even lower. Their radio orphan afterglow
will probably still be detectable in a deep search. However, its
variability time scale is short, about a week, so the search should
be done promptly.

Our results show the great potential of  $1.4$ GHz radio transient
observations at the sub-mJy level for the detection of ns$^2$
mergers. On the observational side these predictions provide an
excellent motivation for carrying out a whole sky sub-mJy survey
using the EVLA or  other upcoming radio telescopes. The main source
of contamination in such surveys would be radio supernovae and those
could be distinguished from compact binary mergers by their optical
signal, optically thick spectrum and other characteristic
properties.

{ The IR-UV light curves expected from ``macronovae'',
supernova-like events powered by the radioactive decay within the
ejecta,  depends on the total mass ejected and on the velocity
distribution.  We present a general method for performing such
calculations. For each of the 23 simulations that cover the binary
parameter space we follow a large number of ejecta trajectories with
a state-of-the-art nuclear reaction network. All trajectories show
very similar radioactive heating histories which were fitted in
\cite{korobkin12}, see their equation~(12). A large uncertainty,
though, comes from the poorly known outflow opacity. If we adopt the
value that has been discussed in some detail in \cite{Metzger+10}, we find
that ns$^2$ [nsbh] mergers peak at about $5 \times 10^{41}$
[$10^{42}$] erg/s, corresponding to absolute bolometric magnitudes
of -15 [-16]. The magnitude in a given observed optical band is
probably fainter by a magnitude or two. Such events can be detected
by current blind surveys like PTF up to a distance of about 150
[300] Mpc and by LSST up to 0.8 [1.5] Gpc. The short duration of
these events, about 0.4 [0.7] days, may pose problem as it would
require very short cadence surveys. Factoring in these limits we
expect for the canonical rate of 300 Gpc$^{-3}$ yr$^{-1}$ ns$^2$
mergers a detection of 0.01 macronovae per year by PTF and 100 per
year by the LSST.  If the opacity is higher than  the value
suggested by \cite{Metzger+10} then the peak time is delayed 
while the peak luminosity drops, making the detection of macronova 
light even harder.

Before concluding we address the relation between mergers the associated radio flares and the short GRBs.  While it has not been confirmed it is possible, and maybe even likely, that compact binary mergers are the origin of short GRBs. For that reason we compared ``orpan afterglowsÓ with radio flares and demonstrated that radio flares are expected to be brighter. It has been claimed \cite{MetzgerBerger12} that sGRBs arise in large distances from the host galaxies regions and that low density is inferred from modeling of their afterglow. Hence they suggest that if the association of mergers and GRBs is correct this implies that mergers take place in low-density regions and as such their radio flares will be undetectable. However, we know that regardless of the question whether sGRBs are associated with mergers or not, the compact binaries that have been observed in our galaxy are in the galactic disk, namely in high ISM density. Merger rate estimates based on these binaries \citep{KalEtal04,KalEtal04a} suggest a comparable rate of events to the beaming corrected rate inferred from sGRBs. This implies that, while there may (or may not) be a merger population at low-density environment, there must be (regardless of the whether there is a connection to sGRBs) a large population of the mergers that take place in Milky Way ISM-like density.

We have presented here the detailed methodology for calculation of
the two more robust EM transients that should accompany compact
binary mergers - radio flares that arise from the interaction of sub-
to mildly relativistic outflows with the surrounding matter and
macronovae that arise from the radioactive decay of the neutron star
matter.  We have obtained these estimates from realistic (employing
a slew of microphysics ingredients) though Newtonian merger
simulations. The major sources of uncertainty in our estimates for
radio flares are the fraction of mildly relativistic ejecta (which
can only increase the emission) and the surrounding matter density.
For the IR-UV macronovae the main uncertainties are the
radioactive energy source within the neutron star matter and its
opacity. We find that the prospects for detection of both radio
flares and the IR-UV macronova are promising by  intensive
follow up searches following GW signals. Radio flares that last a
few months to years have the advantage that they don't require a
rapid followup and that the background sky contamination is rather
low. Macronovae are more challenging, as they require a large field
of view followup within a very short time frame of days or even less
and they need to be identified in the crowded optical transient sky.
Their advantage is that their emission is independent
of the circum-merger environment.\\

\noindent{\bf Acknowledgements}\\
We acknowledge the use of the visualization software SPLASH
developed by Daniel Price (2007). The presented simulations  were
performed on the facilities of the H\"ochstleistungsrechenzentrum
Nord (HLRN).  S.R.'s research  was supported by DFG grant RO-3399,
AOBJ-584282. T.P.'s research was supported by an Advanced ERC grant
and by the Israeli center for Excellence for High Energy
Astrophysics. E.N.'s research was supported by the Israel Science
Foundation (grant No. 174/08) and by an IRG.

\hyphenation{Post-Script Sprin-ger}

\end{document}